\documentclass{jpp}
\usepackage{graphicx,natbib}
\usepackage{epstopdf, epsfig}
\usepackage{amsfonts,amsmath,amssymb}
\usepackage[pdftex,colorlinks,citecolor=blue]{hyperref}
\usepackage{bm}
\usepackage{xcolor}
\usepackage[]{cleveref}
\crefname{section}{\S\!}{\S\S\!}
\crefname{appendix}{\S\!}{\S\S\!}
\crefname{equation}{Eq.}{Eqs.}
\Crefname{equation}{Equation}{Equations}
\crefname{figure}{Fig.}{Figs.}
\Crefname{figure}{Figure}{Figures}
\usepackage{mathtools}
\usepackage{cuted}
\newcommand{\mockalph}[1]{}	
\usepackage{mathtools}
\DeclarePairedDelimiter\abs{\lvert}{\rvert}

\newcommand\revchng[1]{{#1}}
\newcommand\revchngtwo[1]{#1}

\newcommand{\sigf}{\sigma_{\varepsilon}}
\newcommand{\vph}{{v}_{\rm ph}}
\newcommand{\kstar}{k_{\perp}^{*}}
\newcommand{\edprp}{\varepsilon^{\rm diss}_{\perp}}
\newcommand{\edprpH}{\varepsilon^{\rm diss}_{\mathcal{H},\perp}}
\newcommand{\ed}{\varepsilon^{\rm diss}}
\newcommand{\edprl}{\varepsilon^{\rm diss}_{z}}
\newcommand{\nucrit}{\nu_{6\perp}^{\rm crit}}
\newcommand{\kcrit}{k_{\perp}^{\rm crit}}

\ifCUPmtlplainloaded \else
  \checkfont{msam10}
  \iffontfound
    \IfFileExists{amssymb.sty}
      {\typeout{^^JFound AMS Symbol fonts on the system, using the
                'amssymb' package.^^J}%
       \usepackage{amssymb}%
         \let\leq=\leqslant
         
      }{}
  \fi
\fi

\ifCUPmtlplainloaded \else
  \IfFileExists{amsbsy.sty}
    {\typeout{^^JFound the 'amsbsy' package on the system, using it.^^J}%
     \usepackage{amsbsy}}
    {\providecommand\boldsymbol[1]{\mbox{\boldmath $##1$}}}
\fi

\title{On the violation of the zeroth law of turbulence in space plasmas}

\author[R.~Meyrand and others]%
{R.~Meyrand$^{1}$,
 \thanks{Email address for correspondence: romain.meyrand@otago.ac.nz},
  J.~Squire$^{1}$,
 A.~A.~Schekochihin$^{2,3}$,
and W.~Dorland$^{4}$}

\affiliation{$^1$Department of Physics, University of Otago, 730 Cumberland St., Dunedin 9016, New Zealand\\[\affilskip]
$^2$The Rudolf Peierls Centre for Theoretical Physics, University of Oxford, Clarendon Laboratory, Parks Road, Oxford, OX1 3PU, UK\\[\affilskip]
$^3$Merton College, Merton Street, Oxford OX1 4JD, UK\\[\affilskip]
$^4$Department of Physics, University of Maryland, College Park, MD 20742, USA}

\pubyear{2018}
\volume{}
\pagerange{}
\date{?; revised ?; accepted ?. - To be entered by editorial office}

\begin{document}
\maketitle

\begin{abstract}
The zeroth law of turbulence states that, for fixed energy input  into large-scale 
motions, the statistical steady state of a turbulent system is independent of  microphysical  
dissipation properties. This behavior, which is fundamental to nearly all fluid-like systems from industrial processes to galaxies, occurs  
because  nonlinear processes generate smaller and smaller scales in the flow, until  the dissipation---no matter how
small---can thermalise the energy input. Using direct numerical simulations and theoretical arguments, we show that in 
strongly magnetised plasma turbulence such as that recently observed by the  Parker Solar Probe (PSP) spacecraft, 
the zeroth law is routinely violated. Namely, when such turbulence is ``imbalanced''---when the large-scale energy input
is dominated by Alfv\'enic perturbations propagating in one direction (the most common situation in space plasmas)---nonlinear conservation laws imply 
the existence of a ``barrier'' at scales near the ion gyroradius. This causes energy to  build up over time at large scales. 
The resulting magnetic-energy spectra bear a strong resemblance to those observed in situ, exhibiting
a sharp, steep kinetic transition range above and around the ion-Larmor scale, with flattening at yet smaller scales. \revchng{The effect thus offers a possible
solution} to the decade-long puzzle of the position and variability of ion-kinetic spectral breaks in plasma turbulence.
The existence of the ``barrier''  also suggests that how a plasma is forced at large scales (the imbalance) may have a crucial influence on thermodynamic properties such 
as the ion-to-electron heating ratio.
\end{abstract}



\section{Introduction}
In his celebrated 1850 manuscript ``The Mechanical Equivalent of Heat'' \citep{Joule1850}, James Prescott Joule described how
a liquid stirred by a falling mass would heat  up by a well-defined, fixed amount, thus demonstrating the equivalence of mechanical work and heat. 
Though less well appreciated,  Joule's study also revealed another,  similarly intriguing law of nature: 
 his series of experiments, which  measured the work done by different masses on either water or mercury at \revchng{high ($\gtrsim\!10^5$) Reynolds number}, showed that
the damping rate of the liquid's kinetic energy must be proportional to its velocity, rather than  its viscosity. 
This is despite the fact that it is the viscosity that is ultimately responsible for the conversion of work to heat.
The general principle, which  has subsequently become known as the ``zeroth law of turbulence,'' 
states that the dissipation rate of a \revchng{high-Reynolds-number} turbulent flow under fixed large-scale conditions is independent of the value or mechanism of
the microphysical energy dissipation (e.g., the viscosity). This distinctive property arises because turbulence nonlinearly generates motions at successively smaller
 scales, always reaching the scale where viscous effects become large, no matter how small the viscosity itself. 

Collisionless plasmas, although far more complex than water and mercury as used in Joule's experiments, are generally assumed to satisfy the 
zeroth law.
Energy injected into smooth, large-scale fluctuations in position and velocity space (phase space)---for example, Alfv\'enic perturbations
emitted from the Sun's corona---must make its way (linearly or nonlinearly) towards 
small scales before it can be converted to heat. If this is not possible---if the zeroth law is violated---the injected energy
will not be efficiently thermalised, instead building up over time in large-scale motions and magnetic fields. An inability of the system 
to transfer energy to small scales thus has a dramatic impact on the large-scale behavior of the plasma. 
In this paper, we argue that,
counter to the assumptions of much previous work,
the zeroth law can be violated strongly in magnetised (Alfv\'enic) plasma turbulence such as that observed in the solar wind. 
The effect, which occurs when the turbulence is ``imbalanced'' (i.e., when the energies of forward and backward propagating fluctuations differ),
arises because both energy and a ``generalised helicity'' \revchng{(see \cref{eq: helicity})} are nonlinearly conserved in strongly magnetised (low-beta) collisionless plasmas.
At scales above the ion gyroradius $\rho_{i}$, the generalised helicity is the magnetohydrodynamic (MHD) cross-helicity and naturally undergoes a forward cascade (nonlinear energy transfer to small scales);  
at scales below $\rho_{i}$, the generalised helicity becomes magnetic helicity and naturally undergoes an inverse cascade (nonlinear transfer to larger scales; \citealp{Cho2011}). 
The collision of the two cascades creates a ``helicity barrier'': it stops the system from dissipating
injected energy through nonlinear transfer to smaller spatial scales. 

\begin{figure}
\centering
\includegraphics[width=\columnwidth]{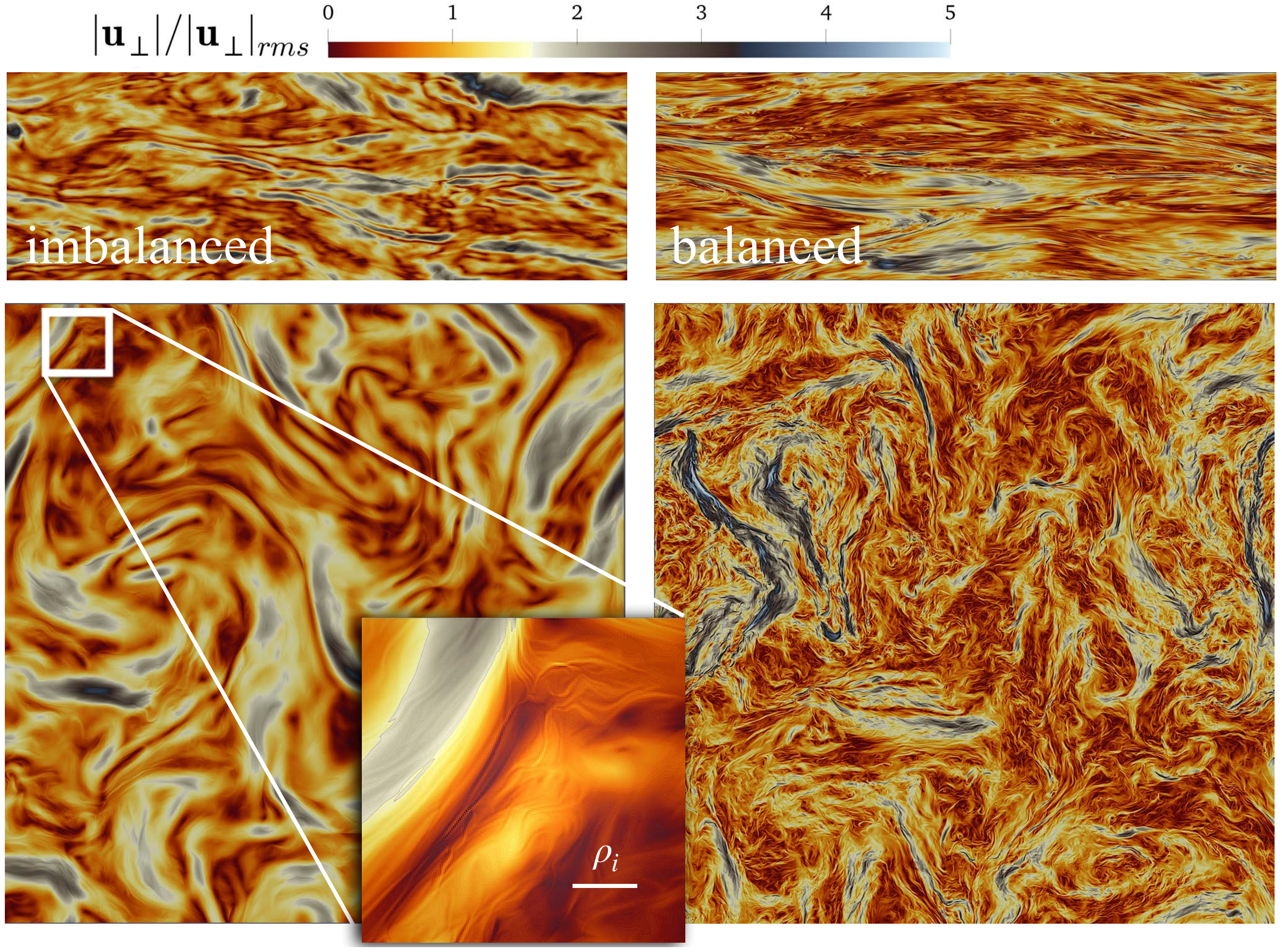}
\caption{The spatial structure of  the perpendicular electron flow $\bm{u}_{\perp}$, or equivalently, the perpendicular electric field $\bm{E}_{\perp}=-\nabla_{\perp}\varphi$ \revchng{[see \crefrange{eq: FLRMHD A}{eq: FLRMHD phi}]}. We compare imbalanced and balanced turbulence in the left and right panels, respectively. Top panels show a parallel $(x,z)$ slice ($\bm{B}_0=B_{0}\hat{\bm{z}}$ left to right), bottom panels show a perpendicular $(x,y)$
slice ($\bm{B}_0$ out of the page). The dramatic dependence on imbalance arises because  imbalanced turbulence is afflicted by the ``helicity barrier'': at a nonuniversal scale $\kstar\rho_i\lesssim 1$ most of the energy cascade of the dominant component ($E^+$) cannot proceed to smaller scales, violating the zeroth law of turbulence. The resulting sharp break in the spectrum is shown in Fig.~\ref{fig:spectra}, and is followed by the re-emergence 
of a cascade at yet smaller scales (see zoomed region of left-hand panel). These simulations 
have a resolution of $2048^3$ and are initialised by refining the $256^3$ simulations
of Figs.~\ref{fig:energy}--\ref{fig:fluxes}, starting at $t\approx18\tau_A$.
}\label{fig:picture}
\end{figure}

\begin{figure}
\centering
\includegraphics[width=0.7\columnwidth]{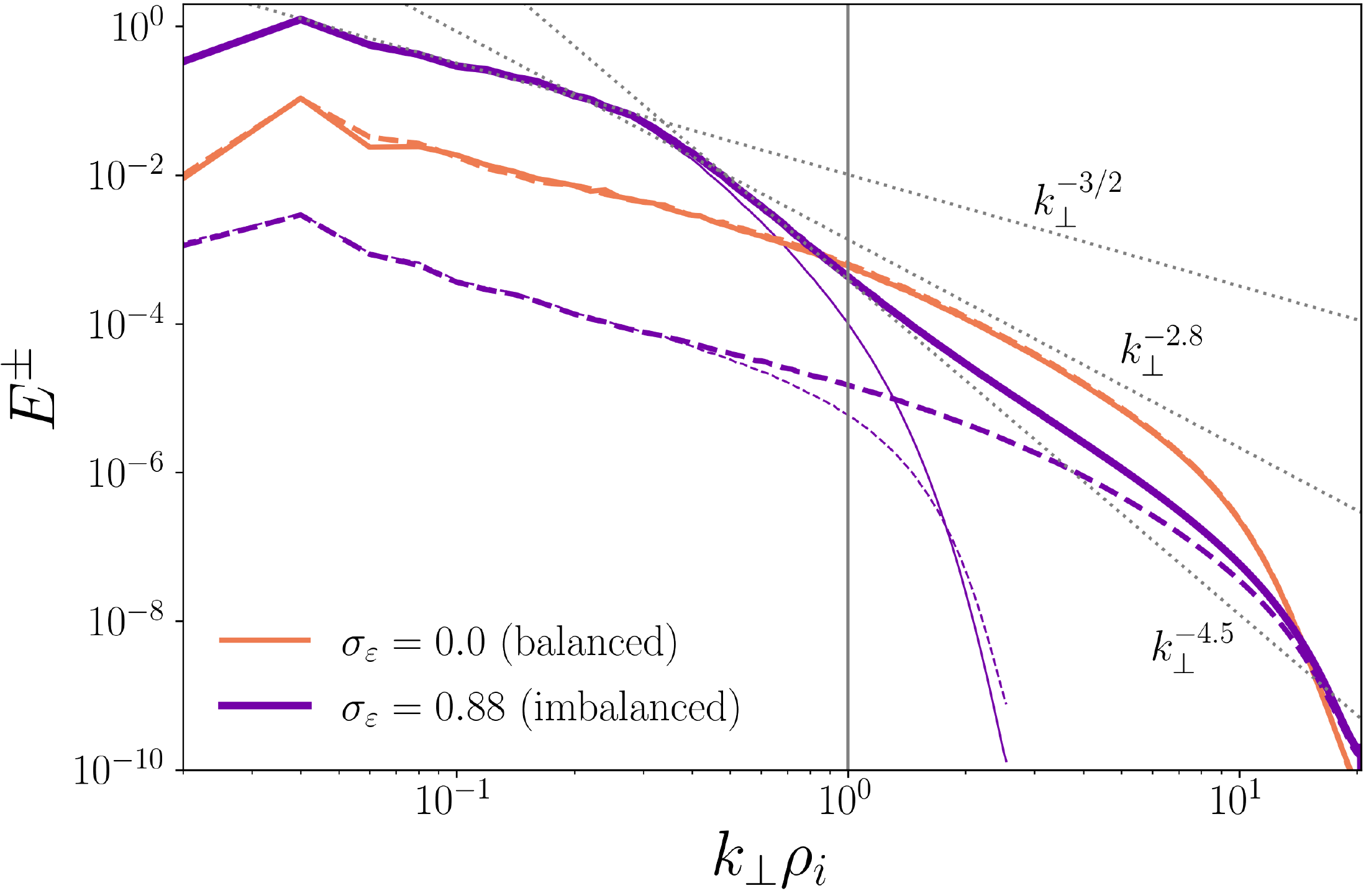}
\caption{Energy spectra for the simulations pictured in Fig.~\ref{fig:picture}. Purple and orange lines  show imbalanced and balanced turbulence, respectively, while solid and dashed lines show the dominant ($E^+$) and subdominant ($E^-$) energies, respectively \revchng{[see Eq.~\eqref{eq: energy}]}. Thin lines show the spectra of the $256^3$ imbalanced simulation at the same time and parameters (see Fig.~\ref{fig:time spectra}),  emphasising
the re-emergence of a kinetic-Alfv\'en-wave cascade ($\sim\!k^{-2.8}$) at small scales in imbalanced 
turbulence, if a sufficient range of scales is available. 
The resulting  double-kinked spectrum strongly  resembles those
observed in the solar wind \citep{Sahraoui2009,Bowen2020a}. As far as we know, this is the first 
time such spectra have been reproduced in a numerical simulation.  }
\label{fig:spectra}
\end{figure}

The resulting turbulence, which we illustrate in Figs.~\ref{fig:picture} and \ref{fig:spectra},
bears a strong resemblance to recent measurements from the Parker Solar Probe (PSP) spacecraft and others. While balanced turbulence shows the expected transition from Alfv\'enic to 
kinetic-Alfv\'en-wave (KAW) turbulence at $\rho_i$ scales  \citep{Howes2008a,Schekochihin2009}, in imbalanced turbulence (purple lines in 
Fig.~\ref{fig:spectra}), the ion-kinetic transition, which is instead controlled by the helicity barrier, is both much sharper and occurs at a larger scale. 
The break in the spectrum is dramatic, with a very steep spectral slope in the transition range, causing manifest differences in the turbulent flow structure compared to balanced turbulence (Fig.~\ref{fig:picture}). Despite the cascade barrier, the energy exhibits a standard $\sim\!k^{-3/2}$ spectrum \citep{Boldyrev2006} above the transition.\footnote{The range  in which it is observed here is not wide enough to distinguish between $k^{-3/2}$ \citep{Maron2001,Boldyrev2006,Perez2012} and $k^{-5/3}$ \citep{Goldreich1995,Beresnyak2014}, but this RMHD-range scaling is not the point of this work. We will compare to $k^{-3/2}$ where necessary because it is well motivated in balanced turbulence and supported by observations \citep{Chen2020}.}
 At yet smaller scales, \revchng{a spectral flattening (approaching the $\sim\!k^{-2.8}$ spectrum expected for KAW turbulence}; \citealp{Schekochihin2009,Alexandrova2009,Alexandrova2012,Boldyrev2013}) is observed due to small leakage through the barrier \revchng{(see \cref{sub: numerical results})}.
The behavior matches observations of near-Sun imbalanced turbulence from PSP, which often show clear spectral breaks significantly above the ion-Larmor scale, with a nonuniversal 
spectrum between the break and a flatter spectrum at yet smaller scales \citep{Bowen2020a,Duan2021}. 
In our theory, \revchng{which differs from previous phenomenologies that assume either an enhanced cascade rate or energy dissipation around $\rho_{i}$ scales \citep[e.g.,][]{Voitenko2016,Mallet2017},} the spectral break occurs around the scale at which the helicity barrier halts the energy flux, 
and this barrier moves to larger scales as the outer-scale energy grows with time.  Final saturation, which occurs only after 
many Alfv\'en crossing times and depends on  simulation resolution,  relies on 
fluctuations reaching large amplitudes and dissipating through  nonuniversal (and, in our simulations, artificial) means. This suggests 
that observed  turbulent cascades in the solar wind may not be in a saturated state where energy input balances dissipation. It \revchng{may} also
 explain the observed non-universality of the break scale and of the sub-break spectral scaling. 

\revchng{The remainder of the paper is organised as follows. Section~\ref{sec: theory} provides  the theoretical framework our study, starting with 
the minimal ``Finite-Larmor-Radius MHD'' model (\cref{sub: flr-mhd description}) used for theoretical arguments and simulations throughout this work.
This is followed by a brief overview of imbalanced turbulence (\cref{sub: imbalanced turbulence}) before 
the presentation of our main theoretical result -- a simple proof that energy and generalised helicity cannot  both cascade
to arbitrarily small (sub-$\rho_{i}$) perpendicular scales, thus causing violation of the zeroth law (\cref{sub: helicity barrier}). Numerical 
results, including the details of the methods used to produce  Figs.~\ref{fig:picture} and \ref{fig:spectra}, are covered 
in \cref{sec: numerical experiments}. We start with numerical demonstrations of the zeroth-law violation in both 
the perpendicular and parallel directions  (\cref{fig:zeroth law}), before considering in more detail the properties of
turbulence that is affected by the helicity barrier (\cref{subsub: the effect of the barrier,subsub: ion spectral break}). Finally,
we explore the possible consequences of our results for  space plasmas in \cref{sec: discussion}. We 
consider the potential impact of other plasma effects that are not contained in our model, followed by a qualitative
discussion of how 
our predictions compare to in-situ observations of solar-wind  turbulence (\cref{sub: implications for the solar wind}).}


\section{Theoretical Framework}\label{sec: theory}
Before continuing, we define the following symbols, with $\alpha$ signifying species (either ions, $\alpha=i$, or electrons, $\alpha=e$):  $n_{0\alpha}$ is the background density;  $T_{0\alpha}$ is the background temperature and $\tau=T_{0i}/T_{0e}$; $\bm{B}$ is the magnetic field, with $\bm{B}_{0}=B_{0}\hat{\bm{z}}$ the background; $\beta_{\alpha} = 8\pi n_{0\alpha}T_{0s}/B_{0}^{2}$ is the ratio of thermal to magnetic energy; $m_{\alpha}$ is the particle mass; $q_{\alpha}$ is the particle charge with $q_{e}=-e$ and $q_{i}=Ze$;  $\Omega_{\alpha}=|q_{\alpha}|B_{0}/m_{\alpha}c$ is the gyroradius; $\rho_{\alpha}=c\sqrt{2m_{\alpha}T_{0\alpha}}/|q_{\alpha}|B_{0}$ is the gyroradius; 
  $d_{\alpha}=\rho_{\alpha}/\sqrt{\beta_{\alpha}}$ is the skin depth;  $c$ is the speed of light; and $v_{A}=B_{0}/\sqrt{4\pi n_{0i}m_{i}}$ is the Alfv\'en speed.
  
In order to elucidate the key physical processes involved in this highly complex problem,
our approach is to use the simplest plasma model that  meets two important requirements:
(i) it can be formally (asymptotically) derived in a physically relevant limit, which allows us to  evaluate critically the plasma regimes in which our results 
remain valid; and (ii) 
 it remains valid for perpendicular scales both above and below the $\rho_{i}$ scale, which 
 is clearly a necessity for a study of the ion-kinetic transition. 
The minimal model of \emph{Finite-Larmor-Radius MHD (FLR-MHD)} described below meets these requirements \citep{Passot2018,Schekochihin2019}, while avoiding the serious complexity of solving  kinetic equations in phase space. It is formally 
valid for  low-frequency Alfv\'enic fluctuations in a  $\beta_{e}\ll1$ plasma, at perpendicular scales above 
 $d_{e}$ and  $\rho_{e}$. Because $\rho_{e}\ll d_{e}$ at $\beta_{e}\ll1$ and $d_{e}/\rho_{i}=(m_{e}/m_{i})^{1/2}/\sqrt{\beta_{i}}$ in a neutral plasma, so long as $\beta_{i}>m_{e}/m_{i}$ and $\beta_{i}\sim\beta_{e}$, 
FLR-MHD provides a valid description of the ion-kinetic transition. The low-$\beta$ assumption
is well satisfied in many astrophysical and space plasmas, including in the solar corona and the near-Sun  solar wind \citep{Bruno2013}.

\subsection{FLR-MHD Model}\label{sub: flr-mhd description}
FLR-MHD can be self-consistently derived from the Vlasov equation, starting  with the assumptions that all fields (the magnetic field, flow velocity etc.) vary slowly in time compared to the  ion-cyclotron frequency,  that there is a strong background magnetic field, and that the
correlation length $l_{\|}$ of a perturbation in the field-parallel direction  is much larger than its field-perpendicular correlation length $l_{\perp}$ \citep{Schekochihin2009}.   
  The resulting system (gyrokinetics) is still quite complex, and significant further
simplification is possible using an  expansion in $\beta_{e}\sim\beta_{i}\ll1$ \citep{Zocco2011,Schekochihin2019}. In this case, the ion-thermal speed
is small compared to the Alfv\'en speed, implying there is minimal coupling between
perpendicular (Alfv\'enic) motions and ion-compressive (kinetic) degrees of freedom, even for ion-Larmor-scale fluctuations \citep{Schekochihin2019}. 
This means that energy injected into Alfv\'enic motions at the largest scales ($l_{\perp}\gg\rho_{i}$) cannot 
directly heat ions (within the low-frequency approximation), allowing the formulation
of a simple closed set of fluid equations (i.e., equations in 3-D space) to describe  the Alfv\'enic component of the 
turbulence both above and below the $\rho_{i}$ scale.\footnote{Compressive fluctuations,
which cascade  passively  to $\rho_{i}$ scales,  where they likely heat the ions through nonlinear phase mixing  \citep{Meyrand2019,Kawazura2020}, 
can modify the equations around $\rho_{i}$ scales by changing the relationship between $\delta n_e$ and $\varphi$ (although they cannot exchange energy with  Alfv\'enic fluctuations; \citealp{Schekochihin2019}). We are effectively assuming throughout this work that the energy in the Alfv\'enic cascade dominates over that in a  compressive cascade, which is (mostly) well justified in the solar wind \citep{Chen2016a}.} 
These are the FLR-MHD equations. We note that the assumption $l_{\perp}\ll l_{\|}$, 
which is well tested in the solar wind \citep{Chen2016a}, is satisfied 
in  standard  magnetised plasma turbulence phenomenologies \citep{Goldreich1995,Boldyrev2006,Schekochihin2020}. The key idea 
is that  of a ``critical balance'' between linear and nonlinear times  at all scales, which  leads to the estimate $l_{\|}\sim l_{\perp}^{1/2}\gg l_{\perp}$ \citep{Boldyrev2006,Mallet2017a}. 
At electron-skin-depth 
scales ($l_{\perp}\sim d_{e}$) where the magnetic field is no longer frozen into the electron flow, FLR-MHD breaks down due to coupling to the electron distribution function. 
Although a model 
exists to  capture this transition accurately \citep{Zocco2011}, its additional complexity 
is unnecessary for describing the ion-kinetic transition of interest here. 
We thus focus on scales above $d_{e}$, which also implies $\beta_{i}>m_{e}/m_{i}$ so that $d_{e}<\rho_{i}$.

The FLR-MHD equations are
\begin{gather}
\left(\dfrac{\partial}{\partial t} +\bm{u}_{\perp}\cdot\nabla_\perp\right)\dfrac{\delta n_{e}}{n_{0e}}  = 
-\frac{c}{4\pi e n_{0e}}   \left(\dfrac{\partial}{\partial z} + \bm{b}_{\perp}\cdot\nabla_\perp\right) \nabla_\perp^2A_\parallel + \mathcal{D}_{6\nu}\dfrac{\delta n_{e}}{n_{0e}}, \label{eq: FLRMHD n}\\
\left(\dfrac{\partial}{\partial t} +\bm{u}_{\perp}\cdot\nabla_\perp\right)A_{\parallel} =-c\dfrac{\partial \varphi}{\partial z}+\dfrac{cT_{0e}}{e}
  \left(\dfrac{\partial}{\partial z} + \bm{b}_{\perp}\cdot\nabla_\perp\right) \dfrac{\delta n_{e}}{n_{0e}}  + \mathcal{D}_{6\nu}A_\|,\label{eq: FLRMHD A}\\
\dfrac{\delta n_{e}}{n_{0e}}=-\dfrac{Z}{\tau}\left(1-\hat{\Gamma}_0\right)\dfrac{e\varphi}{T_{0e}}\label{eq: FLRMHD phi},
\end{gather}
  where $\delta n_{e}/n_{0e}=\delta n_{i}/n_{0i}$ is the perturbed electron (and, by quasi-neutrality, ion) density, $A_{\|}$ is
  the $\hat{\bm{z}}$ component of the vector potential, $\varphi$ is the electrostatic potential,
$\bm{u}_{\perp}=c\,B_{0}^{-1}\hat{\bm{z}}\times\nabla_\perp\varphi$ is the perpendicular $\bm{E}\times\bm{B}$ (electron) flow, and $\bm{b}_{\perp}=-B_{0}^{-1}\hat{\bm{z}}\times\nabla_\perp A_{\|}$ 
is the perturbation of the magnetic field's direction.   The gyrokinetic Poisson operator $1-\hat{\Gamma}_0= 1-I_{0}(\alpha)e^{-\alpha}$, with $\alpha=-\rho_{i}^{2}\nabla_{\perp}^{2}/2$ and $I_{0}$ the modified Bessel function, becomes $1-\hat{\Gamma}_0\approx - \rho_{i}^{2}\nabla_{\perp}^{2}/2$ for fluctuations with $k_{\perp}\rho_{i}\ll1$, and $1-\hat{\Gamma}_0\approx 1$ for fluctuations with $k_{\perp}\rho_{i}\gg1$.
In the former limit, the FLR-MHD system becomes the well-known Reduced MHD (RMHD) model \citep{Strauss1976}, in the latter 
it becomes the Electron RMHD model \citep{Schekochihin2009,Boldyrev2013}.
\revchng{The hyper-diffusion operator, $\mathcal{D}_{6\nu}=\nu_{6\perp}\nabla_\perp^6+\nu_{6z}\nabla_z^6$, is necessary in order to dissipate energy above the grid scale in our numerical simulations, but
is not intended to model a specific physical process.}

\subsection{Imbalanced Alfv\'enic Turbulence}\label{sub: imbalanced turbulence}

A linearization of Eqs.~\eqref{eq: FLRMHD n}--\eqref{eq: FLRMHD phi}, assuming a sinusoidal spatial dependence with wavenumber $\bm{k}=k_{\perp}\hat{\bm{x}}+k_{z}\hat{\bm{z}}$, yields forward and backward propagating modes of frequency $\omega=\pm k_{z} \vph(k_{\perp})v_{A}$, where 
\begin{equation}
{\vph(k_{\perp})}= \frac{k_\perp \rho_i }{\sqrt{2}} 
\left(\dfrac{1}{1-\hat{\Gamma}_{0}}+\dfrac{Z}{\tau}\right)^{1/2}\approx
\begin{cases} 1 &k_\perp\rho_i\ll1 ,\\
\left({\dfrac{1}{2}+\dfrac{Z}{2\tau}}\right)^{1/2} {k_\perp \rho_i} &k_\perp\rho_i\gg1 .
\end{cases}
\end{equation}
FLR-MHD thus recovers shear-Alfv\'en waves when $k_{\perp}\rho_{i}\ll1$ and (low-$\beta$) kinetic Alfv\'en waves (KAWs) when $k_{\perp}\rho_{i}\gg1$.
The eigenfunctions of these linear modes, known as the  generalised Els\"asser potentials, will provide 
a useful basis for intuitive discussion of the nonlinear problem and turbulence. At wavenumber $\bm{k}$, these
are \begin{equation}
\Theta_{\bm{k}}^{\pm} =-\Omega_{i}\frac{\vph(k_{\perp})}{ k_{\perp}^{2}} \frac{\delta n_{e}}{n_{0e}} \mp  \frac{A_{\|}}{\sqrt{4\pi m_{i} n_{0i} }}.\label{eq: theta def}
\end{equation}
At large scales $k_{\perp}\rho_{i}\ll1$, they have the property $\hat{\bm{z}}\times \nabla_{\perp}\Theta^{\pm}=\bm{Z}^{\pm}=\bm{u}_{\perp}\pm \bm{B}_{\perp}/\sqrt{4\pi m_{i}n_{0i}}$, where $\bm{Z}^{\pm}$ are  the Els\"asser variables \citep{Elsasser1950}.

The utility of $\Theta^{\pm}$ arises from the fact that 
at large scales (i.e., in the RMHD limit), nonlinear interaction---and thus the turbulent cascade---requires the interaction between $\bm{Z}^{+}$ and $\bm{Z}^{-}$ (equivalently, $\Theta^{+}$ and $\Theta^{-}$).
Thus, the difference in amplitude of $\bm{Z}^{+}$ and $\bm{Z}^{-}$, which is known as the \emph{energy imbalance} and is determined
by the outer-scale forcing of the plasma, has a strong influence on the properties of 
the turbulent cascade. We will quantify it in the standard way with \begin{equation}
\sigma_{c}= \frac{\int d^{3}\bm{x}\,(|\bm{Z}^{+}|^{2}-|\bm{Z}^{-}|^{2})}{\int d^{3}\bm{x}\,(|\bm{Z}^{+}|^{2}+|\bm{Z}^{-}|^{2})},\end{equation}
so $\sigma_{c}=\pm1$ if $\bm{Z}^{-}=0$ or $\bm{Z}^{+}=0$.
Although imbalanced RMHD turbulence remains poorly understood \citep{Perez2009,Chandran2008,Beresnyak2009,Lithwick2007,Chandran2019,Schekochihin2020}, 
observations show that solar-wind turbulence is usually imbalanced, particularly in 
near-Sun regions where $|\sigma_{c}|\gtrsim0.9$ \citep{McManus2020}. This occurs because Alfv\'enic perturbations are launched outwards from the corona and only generate an inwards propagating component  due to their interaction with background density and field gradients \citep{Velli1993,Chandran2019}.
Our understanding of plasma turbulence thus remains incomplete without addressing the effect of the imbalance on the flow of energy.

At sub-ion scales ($k_{\perp}\rho_{i}\gg1$), the dispersive nature of  KAWs makes possible nonlinear interactions between 
co-propagating perturbations (e.g., $\Theta^{+}$ with $\Theta^{+}$). 
This implies that the two components can exchange energy and that a turbulent cascade is, in principle, possible with
just one component $\Theta^{\pm}$ \citep{Cho2011,Kim2015,Voitenko2016}.

\subsection{The ``Helicity Barrier''}\label{sub: helicity barrier}

Here we argue that the conservation properties of FLR-MHD imply that a
turbulent flux of energy cannot proceed in the usual way to small scales (where it needs to get to be dissipated)\revchng{, violating the zeroth law. }
We term the barrier in the cascade at scales $l_{\perp}\sim \rho_{i}$, the ``helicity barrier.''

\revchng{
As discussed above, a necessary (though not sufficient) condition for a turbulent system to satisfy the zeroth law is that it
has the ability to transfer energy from the largest scales, where energy is assumed to be injected by external 
processes, to the smallest, where it can be dissipated. The concept can be formalised by the idea
of a turbulent ``flux'' through scale space; if a system is to remain in statistical steady state,   each of its nonlinearly conserved invariants with a source at large scales must have
an associated  flux.\footnote{\revchng{The case
of decaying turbulence can be more subtle; it clearly violates the zeroth law  for some initial 
conditions, even in hydrodynamics or RMHD. Specifically, if the system decays into a large-scale nonlinear solution -- for example,  initially 
imbalanced RMHD turbulence decays into an Els\"asser state (only one of $\bm{Z}^{\pm}$, with no nonlinear interactions; see \citealp{Dobrowolny1980a,Oughton1994,Maron2001,Chen2011a}) -- 
the subsequent (slow) decay of this state will clearly depend on the microphysical dissipation. }} This flux must be constant across all scales until the invariant can be dissipated.
Importantly, this must hold for \emph{all} nonlinear invariants together: if the flux of one invariant is non-constant and/or insufficient to allow its dissipation, 
the invariant will change in time, implying that the system is not in steady state. This concept is familiar in the study of two-dimensional
hydrodynamics, where the dual conservation of energy and enstrophy stops the flux of energy to small scales, causing
both energy and enstrophy to build up in time (in the absence of a large-scale dissipation mechanism).
 The zeroth law can thus be studied in terms of either the dissipation properties, considering how the dissipation 
 rate varies with  microphysical dissipation at fixed large-scale conditions  \citep[e.g.,][]{Pearson2004}, or in terms of the 
 large-scale properties, considering how the statistics of the largest scales  depend on 
 the microphysical dissipation when the rate of energy injection is fixed. 
 In the following, we examine the conservation laws of FLR-MHD and prove that they do not admit 
 a solution with a constant flux of energy to small perpendicular scales when the energy injection is imbalanced at large scales. This implies that large-scale
 flows and fields cannot reach steady state through perpendicular dissipation, violating the zeroth law.\footnote{\revchng{The zeroth law is well 
 known to require relatively large Reynolds numbers to be satisfied \citep{Pearson2004}; indeed, some theories of imbalanced
 RMHD turbulence \citep{Chandran2008,Schekochihin2020} imply a logarithmic dependence of spectral slopes on dissipation parameters,
 technically violating the zeroth law. Numerical simulations have yet to yield a definitive verdict on the validity of such theories \citep{Beresnyak2009,Schekochihin2020}.
The FLR-MHD results below, however, demonstrate a far more brutal zeroth-law violation than these RMHD theories, because they
imply that the system's conserved invariants cannot dissipate. Accordingly, our focus here is not
on subtle details of  Reynolds-number dependence  or whether RMHD can also (modestly) violate the 
zeroth law. In any case, our simulations provide direct comparisons of FLR-MHD to RMHD, showing that they differ markedly 
in their large-scale saturation properties.}}
 }

\subsubsection{Conservation laws of FLR-MHD}
The FLR-MHD system has two nonlinearly conserved quadratic invariants, (free) energy and  (generalised) helicity. 
These are most easily and clearly written 
in terms of the generalised Els\"asser variables. The free energy is
\begin{equation}
E=\dfrac{1}{4}\sum_{\bm{k}}\left( |\,k_{\perp}\Theta_{\bm{k}}^{+}|^{2} + |\,k_{\perp}\Theta_{\bm{k}}^{-}|^{2} \right),\label{eq: energy}
\end{equation} 
which reduces to \begin{equation}
E\approx \frac{1}{4}\int \frac{d^{3}\bm{x}}{V}\,(|\bm{Z}^{+}|^{2}+|\bm{Z}^{-}|^{2})=\frac{1}{2}\int \frac{d^{3}\bm{x}}{V}\,\left(|\bm{u}_{\perp}|^{2}+\frac{|\bm{B_{\perp}}|^{2}}{4\pi n_{0i}m_{i}}\right)
\end{equation}
at large scales. 
The generalised helicity is 
\begin{equation}
\mathcal{H}=\frac{1}{4}\sum_{\bm{k}} \frac{|\,k_{\perp}\Theta_{\bm{k}}^{+}|^{2} - |\,k_{\perp}\Theta_{\bm{k}}^{-}|^{2}}{\vph(k_{\perp})},\label{eq: helicity}
\end{equation} 
which  reduces to the MHD cross-helicity  at $k_{\perp}\rho_{i}\ll1$, $\mathcal{H}\propto\int d^{3}\bm{x}\,\bm{u}_{\perp}\cdot\bm{B}_{\perp}$, and becomes  magnetic helicity at $k_{\perp}\rho_{i}\gg1$,  $\mathcal{H}\propto \int d^{3}\bm{x}\, \delta B_{\|} A_{\|}$\footnote{Here $\delta B_{\|}$ is the magnetic-field
strength perturbation; $\delta B_{\|}\propto \delta n_{e}$ for $k_{\perp}\rho_{i}\gg1$ \citep{Schekochihin2009}.}.
If the $k_{\perp}\rho_{i}\ll1$ motions dominate over the smaller scales, the energy imbalance is $\sigma_{c}\approx \mathcal{H}/E$.
 We also define the $\Theta^{\pm}$ ``energies,'' $E^{\pm}=\sum_{\bm{k}}|\,k_{\perp}\Theta_{\bm{k}}^{\pm}|^{2} /4$, along with perpendicular spectra for $E$, $\mathcal{H}$, and $E^{\pm}$, denoted $E(k_{\perp})$, $E_{\mathcal{H}}(k_{\perp})$, and $E^{\pm}(k_{\perp})$, respectively.

\subsubsection{The inevitability of the helicity barrier}\label{subsub: inevitability of barrier}
Consider the case where energy and helicity are injected at large scales at the rates $\varepsilon$ and $\varepsilon_{\mathcal{H}}$, respectively, 
with  \emph{injection imbalance} $\sigf\equiv|\varepsilon_{\mathcal{H}}|/\varepsilon$.
The conservation laws above tell us that in a statistical steady state, there must be a nonzero  energy flux $\Pi(k_{\perp})$ and helicity flux  $\Pi_{\mathcal{H}}(k_{\perp})$ to small scales where they can be dissipated. If we further assume that (i) energy transfer due to nonlinearity
is significant only for modes with  similar scales (locality), and (ii)  parallel dissipation is small because  eddies are highly elongated along the magnetic field, then $\Pi(k_{\perp})$ and $\Pi_{\mathcal{H}}(k_{\perp})$ must be constant between the forcing and dissipation scales. 
 In the following argument, based on \citet{Alexakis2018},  we assume such a constant-flux solution and find a contradiction, 
 suggesting that this type of solution  is not possible in FLR-MHD when $\sigf\neq0$.
 Fundamentally, the contradiction arises because at large scales $\mathcal{H}$ is the RMHD cross-helicity, which undergoes a forward cascade, while at small
scales $\mathcal{H}$ is magnetic helicity,  which undergoes an inverse cascade \citep{Schekochihin2009,Cho2011,Kim2015,Miloshevich2020,Pouquet2020}.

%
Mathematically, the constant-flux solution 
is \begin{subequations}\begin{gather}
\Pi(k_{\perp})\simeq \varepsilon\simeq \edprp = \nu_{n}\sum_{k_{\perp}}k_{\perp}^{2n} E(k_{\perp}), \\
\Pi_{\mathcal{H}}(k_{\perp})\simeq\varepsilon_{\mathcal{H}}\simeq  \edprpH = \nu_{n}\sum_{k_{\perp}}k_{\perp}^{2n} E_{\mathcal{H}}(k_{\perp}),
\end{gather}\end{subequations}
where $\edprp$ and $\edprpH$ are the energy and helicity dissipation rates 
(we assume  hyper-viscous dissipation of $\delta n_{e}$ and $A_{\|}$ of the form $\nu_{n}k_{\perp}^{2n}$).
This solution satisfies the following inequalities: 
\begin{align}
|\Pi_{\mathcal{H}}(k_{\perp})|&\simeq \nu_{n}\abs[\Big]{\sum_{p_{\perp}=k_{\perp}}^{\infty}p_{\perp}^{2n} E_{\mathcal{H}}(p_{\perp})}\leq \nu_{n} \vph^{-1}(k_{\perp})\abs[\Big]{ \sum_{p_{\perp}=k_{\perp}}^{\infty} p_{\perp}^{2n}  \vph (p_{\perp})E_{\mathcal{H}}(p_{\perp})}\nonumber \\ 
 &
 \leq \vph^{-1}(k_{\perp})\nu_{n} { \sum_{p_{\perp}=k_{\perp}}^{\infty} p_{\perp}^{2n}  E(p_{\perp})}\simeq \vph^{-1}(k_{\perp}) \Pi(k_{\perp}),\label{eq: barrier argument}
\end{align}
where we have used the fact that $\vph(k_{\perp})$ is a monotonically increasing function of $k_{\perp}$, as well as  
 the inequality $\vph(k_{\perp}) |E_{\mathcal{H}}(k_{\perp}) |\leq E(k_{\perp})$   from  Eqs.~\eqref{eq: energy}--\eqref{eq: helicity}.
The ratio of fluxes $|\Pi_{\mathcal{H}}(k_{\perp})|/\Pi(k_{\perp})\simeq \sigf$ must thus satisfy  $\sigf\leq 1/\vph(k_{\perp})$ for all $k_{\perp}$ above the dissipation scales. But $1/\vph(k_{\perp})$ decreases  with $k_{\perp}$
to arbitrarily small values ($\vph\propto k_{\perp}$ at $k_{\perp}\rho_{i}\gg 1$). This suggests that, no matter what the injection imbalance, a cascade that tries to proceed to  small 
scales will at some $k_{\perp}$ violate the inequality \eqref{eq: barrier argument}\footnote{It is worth commenting briefly on the recent work of \citet{Milanese2020}, which has considered a 
similar system with a conserved energy and generalised helicity. In that system, the function $1/\vph(k_{\perp})$ in the generalised helicity
\emph{increases} with $k_{\perp}$ at large $k_{\perp}$, which is the opposite of  our Eq.~\eqref{eq: helicity}. This leads to the phenomenon of ``dynamic phase alignment'', whereby
fluctuations become increasingly correlated at small scales, reducing the 
strength of their nonlinear interaction so as to maintain constant fluxes of energy and helicity.}.  In such a case, the constant-flux solution fails, indicating that
the system is unable to  thermalise energy and helicity input through small-scale dissipation.
We further see that the failure occurs only 
below the  scale where
$ 1/\vph(k_{\perp})\simeq \sigf$; this is around  $k_{\perp}\rho_{i}\simeq1$  for $\sigf\approx 0.7$ but moves to larger scales with increasing $\sigf$. This highlights an interesting difference compared to the well-known inverse energy cascade  of two-dimensional hydrodynamics \citep{Fjortoft1953,Alexakis2018}:
while standard inverse cascades inhibit forward transfer already at the injection scale, 
 helicity must first travel to microphysical ($\rho_{i}$) scales  before it hits the barrier. As a consequence, 
despite FLR effects not  influencing directly the nonlinear interactions at MHD scales, they 
could strongly influence turbulence statistics at those scales by insulating them from the dissipation scales.

\section{Numerical Experiments}\label{sec: numerical experiments}

The argument above suggests that it is not possible to have a constant flux of both energy and 
helicity through the ion-kinetic transition scale. 
It does not, however, 
elucidate  how the system  behaves in the presence of continuous imbalanced injection of energy at large scales.
For this, we turn to numerical simulations.

\subsection{Numerical setup}\label{sub: numerical setup}

We solve Eqs.~\eqref{eq: FLRMHD n}--\eqref{eq: FLRMHD phi} using a modified version of the pseudospectral code TURBO \citep{Teaca2009} in a cubic box $L_{\perp}=L_{z}=L$ with $N_{\perp}^{2}\times N_{z}$ Fourier modes. 
A third-order modified \citet{Williamson1980}  algorithm is used for  time stepping. 
The values of hyper-dissipation coefficients $\nu_{6\perp}$ and $\nu_{6z}$ are chosen based on the numerical  resolution, ensuring that the energy spectrum falls 
off sufficiently rapidly before the resolution cutoff. Fluctuations are stirred at large scales by  added forcing terms ($f^{n_{e}}$ and $f^{A_{\|}}$) in 
Eqs.~\eqref{eq: FLRMHD n}--\eqref{eq: FLRMHD A}. This forcing is confined to $0< k_{\perp}\leq 4\pi/L$ and $\vert k_{z} \vert = 2\pi/L$ and 
takes the form of negative damping ($f^{n_{e}}$ and $f^{A_{\|}}$ proportional to the large-scale modes of $n_{e}$ and $A_{\|}$); this method allows the level of energy and helicity injection ($\varepsilon$ and  $\varepsilon_{\mathcal{H}}$) to be 
controlled exactly, while producing sufficiently chaotic motions to generate turbulence. 
While  $\sigf=\varepsilon_{\mathcal{H}}/\varepsilon$ is thus fixed, the imbalance $\sigma_{c}\approx\mathcal{H}/E$ is determined by the turbulence and evolves in time. 
Initial conditions are random and large-scale with energy $E=10\varepsilon\tau_{A}$, 
where  $\tau_A=L_z/v_A$ is the Alfv\'en crossing time.  
The perpendicular and parallel energy dissipation rates are $\edprp=\nu_{6\perp}\sum_{k_{\perp},k_{z}}k_\perp^6 E(k_\perp,k_z)$ and $\edprl=\nu_{6z}\sum_{k_{\perp},k_{z}}k_z^6 E(k_\perp,k_z)$, where $E(k_\perp,k_z)$ is the 2-D energy spectrum (in steady state, if it exists, we would have $\varepsilon=\ed=\edprp+\edprl$).
Simulations are run across a  range of resolutions up to $N_{\perp}=N_{z}=2048$. For the highest-resolution cases, we use a recursive 
refinement procedure, restarting a lower-resolution case at twice the resolution and running until $\edprp$ converges in time; this 
dramatically reduces the computational cost to enable otherwise unaffordable simulations. All simulations use  $Z=1$ and $\tau=0.5$ (so that the ion-sound radius is equal to $\rho_{i}$).

\begin{figure}
\centering
\includegraphics[width=1.0\columnwidth]{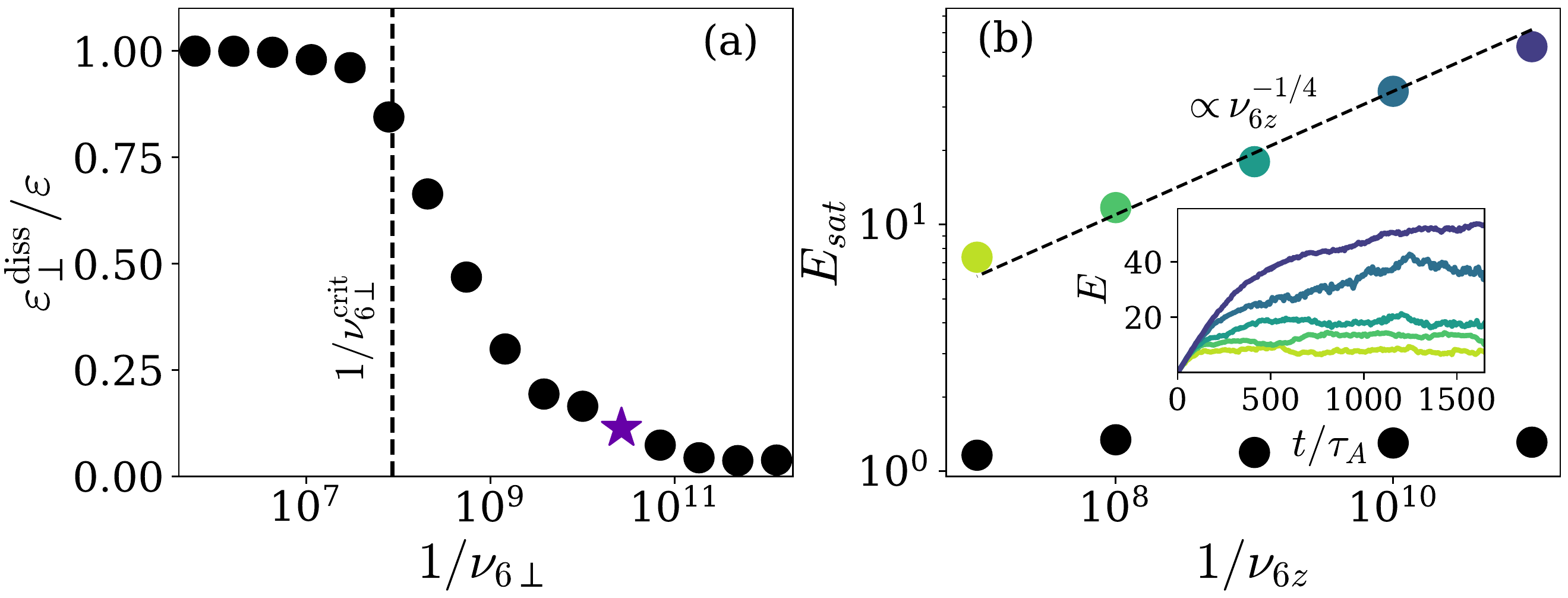}
\caption{ \revchng{Panel (a) illustrates the violation of the zeroth law of turbulence with respect to the perpendicular dissipation. Each point shows $\edprp/\varepsilon$ in the saturated 
state of an FLR-MHD simulation with a different value of $\nu_{6\perp}$. The simulations all have $\sigf=0.88$, $\rho_{i}=0.02L_{\perp}$,  $256\leq N_{\perp}\leq512$, $N_{z}=256$, and are 
initialised from the saturated state of the simulation marked by the purple star and run until they reach steady state. The vertical dashed line shows the critical $\nu_{6\perp}$ at which the RMHD dissipation scale lies above the scale of the helicity barrier (see text), so that simulations with $1/\nu_{6\perp}<1/\nu_{6\perp}^{\rm crit}$ are ignorant of FLR effects (the turbulence is dissipated at larger scales), while those with $1/\nu_{6\perp}>1/\nu_{6\perp}^{\rm crit}$ are not. We see that the helicity barrier halts the perpendicular dissipation causing $\edprp\ll \varepsilon$ at small $\nu_{6\perp}$. } 
In panel (b), coloured points show the saturation energy $E_{\rm sat}$ versus parallel hyper-dissipation $\nu_{6z}$ for $5$ FLR-MHD simulations with $N_\perp=64$, $N_z\leq256$, $\sigf=0.88$, and $\rho_i=0.1L_{\perp}$. Equivalent RMHD simulations ($\rho_i=0$) are shown with black points. The dependence of   $E_{\rm sat}$ on $\nu_{6z}$ at fixed $\varepsilon$ demonstrates
that the helicity barrier causes the violation of the zeroth law of turbulence with respect to the parallel dissipation. The inset shows the time evolution of the energy in each case (colours match those of the points).
}\label{fig:zeroth law}
\end{figure}

\subsection{Results}\label{sub: numerical results}

\begin{figure}
\centering
\includegraphics[width=\columnwidth]{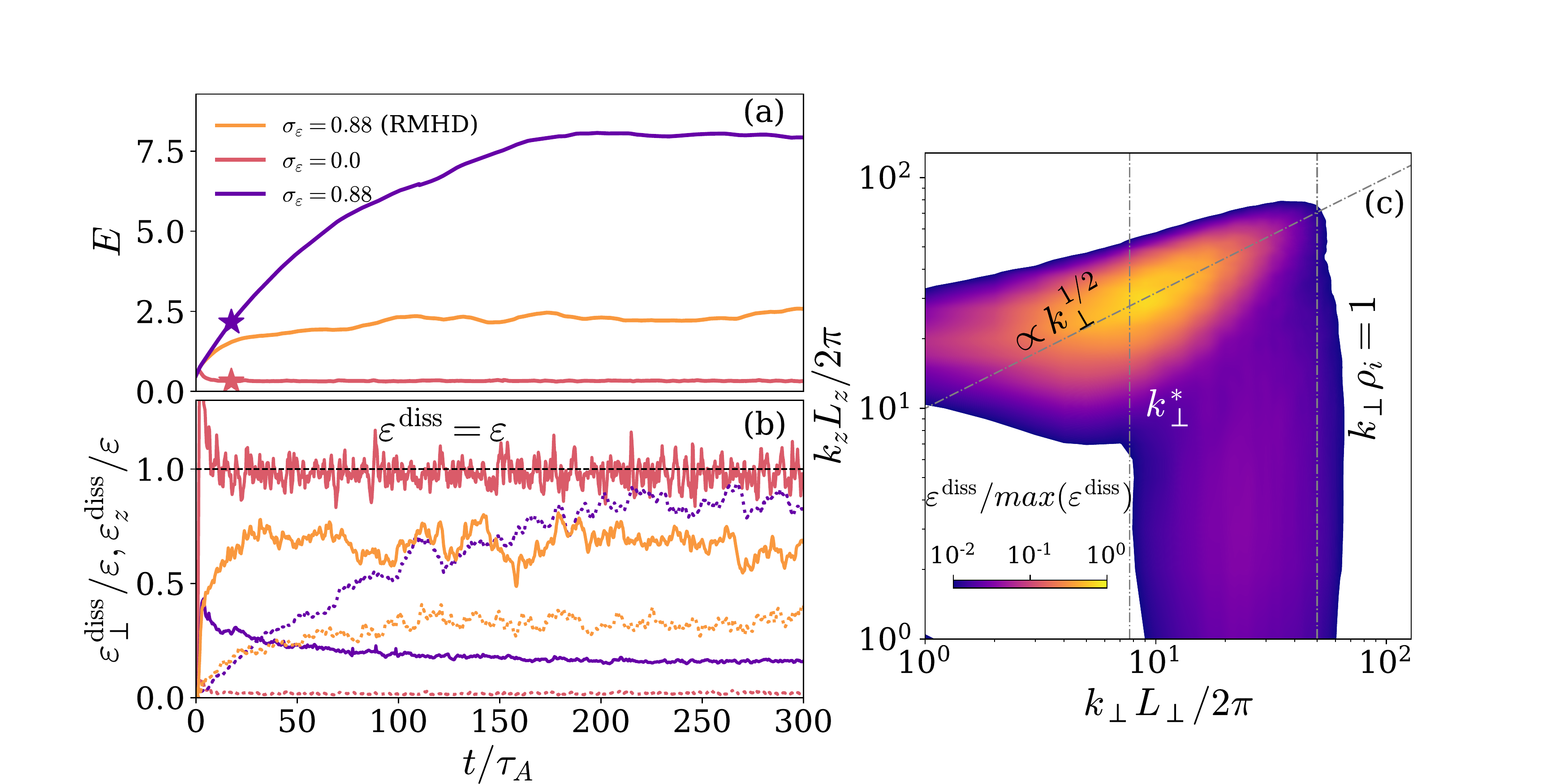}
\caption{Energy and dissipation properties from a set of simulations at resolution $N_\perp=N_z=256$. Panel (a) compares the time evolution of energy in imbalanced FLR-MHD ($\sigf=0.88$, $\rho_i=0.02L$) to balanced FLR-MHD ($\sigf=0$, $\rho_i=0.02L$) and 
imbalanced RMHD ($\sigf=0.88$, $\rho_i=0$). The stars indicate the time from which the higher-resolution simulations of Figs.~\ref{fig:picture}--\ref{fig:spectra} were initialised. 
Panel (b) shows  $\edprp$ (solid lines) and $\edprl$ (dotted lines) for each case, to show that saturation is reached through \emph{parallel} dissipation (unlike in balanced turbulence and in imbalanced RMHD).  Panel (c) shows the ($k_{\perp},k_{z}$) dissipation spectrum in the saturated state of imbalanced FLR-MHD, illustrating
that dissipation occurs primarily at the perpendicular break scale ($\kstar\rho_{i}\simeq0.15$) at high $k_{z}$.}\label{fig:energy}
\end{figure}

\revchng{Figure \ref{fig:zeroth law} demonstrates how imbalanced FLR-MHD turbulence violates the zeroth law of turbulence.
\Cref{fig:zeroth law}(a) shows the normalised perpendicular dissipation rate $\edprp/\varepsilon$ in the saturated state of sixteen FLR-MHD simulations with 
$\sigf=0.88$, $\rho_{i}=0.02L_{\perp}$, and
varying  perpendicular 
hyper-dissipation $\nu_{6\perp}$. The simulations have fixed $\nu_{6z}$, resolution $N_{\perp}=N_{z}=256$ (or $N_{\perp}=512$ for the four lowest-$\nu_{6\perp}$ cases), and were run by restarting from the saturated state of the $1/\nu_{6\perp}\simeq 3\times 10^{10}$ simulation  (denoted by the purple star), which will be discussed in detail below.  We see a precipitous drop in $\edprp$ for  $1/\nu_{6\perp}\gtrsim 10^{8}$, which signifies that the turbulent energy flux cannot proceed to 
small perpendicular scales for small $\nu_{6\perp}$. This is  the signature of the helicity barrier predicted above in \cref{subsub: inevitability of barrier}:
if the scale at which the energy dissipates due to $\nu_{6\perp}$ is such that the inequality \eqref{eq: barrier argument} is violated, the 
constant perpendicular flux solution cannot hold.} 

\revchng{The vertical dashed line in \cref{fig:zeroth law}(a) shows the critical value $\nucrit$ needed to set the perpendicular dissipation scale of RMHD turbulence
equal to the scale $\kcrit\simeq 22.4(2\pi/L_{\perp})$ where $1/\vph(\kcrit)=\sigf=0.88$. We estimate $\nucrit$ from $\nu_{6\perp}k_\perp^6\sim k_\perp Z^-_{k_\perp}$, where $Z^-_{k_\perp}\sim Z^-_{0}(k_\perp L_\perp)^{-1/4}$ is the typical variation in $Z^-$ across scale $k_\perp^{-1}$, and the
outer-scale amplitude $Z^-_{0}$ is taken from saturated state of the similar RMHD simulation shown in \cref{fig:time spectra}.
 If $\nu_{6\perp}>\nucrit$, the cascade can dissipate in the standard way on  hyper-dissipation, setting up an imbalanced RMHD cascade;
if $\nu_{6\perp}<\nucrit$, the inequality \eqref{eq: barrier argument} applies, stopping the cascade before it reaches small perpendicular scales and causing $\edprp\ll\varepsilon$.}

\revchng{This inability of the perpendicular cascade to process  injected energy  into perpendicular dissipation suggests that 
the \emph{parallel}  dissipation must play a role. This is surprising given that parallel dissipation is usually
 neglected in  magnetised-turbulence theories because the increasing elongation of
eddies at smaller scales generally implies $\edprp\gg\edprl$. 
In Fig.~\ref{fig:zeroth law}(b), we show the 
 turbulent-energy saturation amplitude $E_{\rm sat}$ as a function of  the parallel hyper-dissipation $\nu_{6z}$. 
These simulations are again forced with injection imbalance $\sigf=0.88$ and use a lower resolution $N_{\perp}=64$ and $64\leq N_{z}\leq 256$  (with $N_{z}$ chosen as appropriate for each $\nu_{6z}$) because of the long saturation times.
The perpendicular dissipation is fixed and  
 sufficiently small for there to be a helicity barrier. 
We compare  FLR-MHD turbulence with $\rho_{i}=0.1L_{\perp}$ (coloured points) to RMHD  turbulence ($\rho_{i}=0$; black points) to demonstrate the significant role played by FLR effects. 
 The  difference is obvious: FLR-MHD turbulence saturates at much larger amplitudes, which  increase with decreasing dissipation;  larger amplitudes are associated with longer saturation times   
(see inset; cf.~\citealp{Miloshevich2020}).  This dependence of saturation time and large-scale 
properties  on $\nu_{6z}$ shows that imbalanced FLR-MHD turbulence   violates the zeroth law of turbulence with respect to the parallel, 
as well as the perpendicular, dissipation. }

\revchng{The implications of these findings are twofold. First,  in order to saturate, imbalanced FLR-MHD turbulence must access small-scale parallel physics,  escaping
the ordering assumptions ($l_{\|}\gg l_{\perp}$) used to derive the FLR-MHD model in the first place. This suggests that  
detailed properties of the saturated state achieved by this model are not relevant to real physical systems.
Secondly, when the helicity barrier  halts the perpendicular cascade,  the system does not develop a true
parallel cascade that can process the energy and helicity input into parallel dissipation. Rather, its
saturated amplitude depends directly on the microphysical parallel dissipation (the specific $\sim\!\nu_{6z}^{-1/4}$ dependence is explained below),
suggesting that in the limit of vanishing parallel dissipation, the turbulence would fail to saturate completely.}

\revchng{These highly unusual characteristics motivate 
a more thorough exploration of imbalanced FLR-MHD  
turbulence.} Below and in Figs.~\ref{fig:picture}--\ref{fig:spectra},
we present detailed simulation results  to help explain the effect of the helicity barrier 
and its potential relevance to space plasmas. We compare   imbalanced FLR-MHD  to an equally imbalanced RMHD simulation
 and  balanced FLR-MHD, all at the same $\varepsilon$. 
To aid discussion, we break the time evolution into three phases: first, a transient phase during which small-scale motions are produced from the initial conditions; next, a
\emph{pseudo-stationary phase}, which  is the long phase of slow energy growth (seen  in the inset of Fig.~\ref{fig:zeroth law}b)  that occurs due to the helicity barrier; and finally, saturation,
when $\varepsilon\approx\edprp+\edprl$ and $\partial_{t}E\approx0$. During the pseudo-stationary phase and saturation, the helicity barrier creates a sharp 
break in the perpendicular spectrum at a wavenumber that we will denote $\kstar$.

\subsubsection{The effect of the helicity barrier}\label{subsub: the effect of the barrier}
\begin{figure}
\centering
\includegraphics[width=0.64\columnwidth]{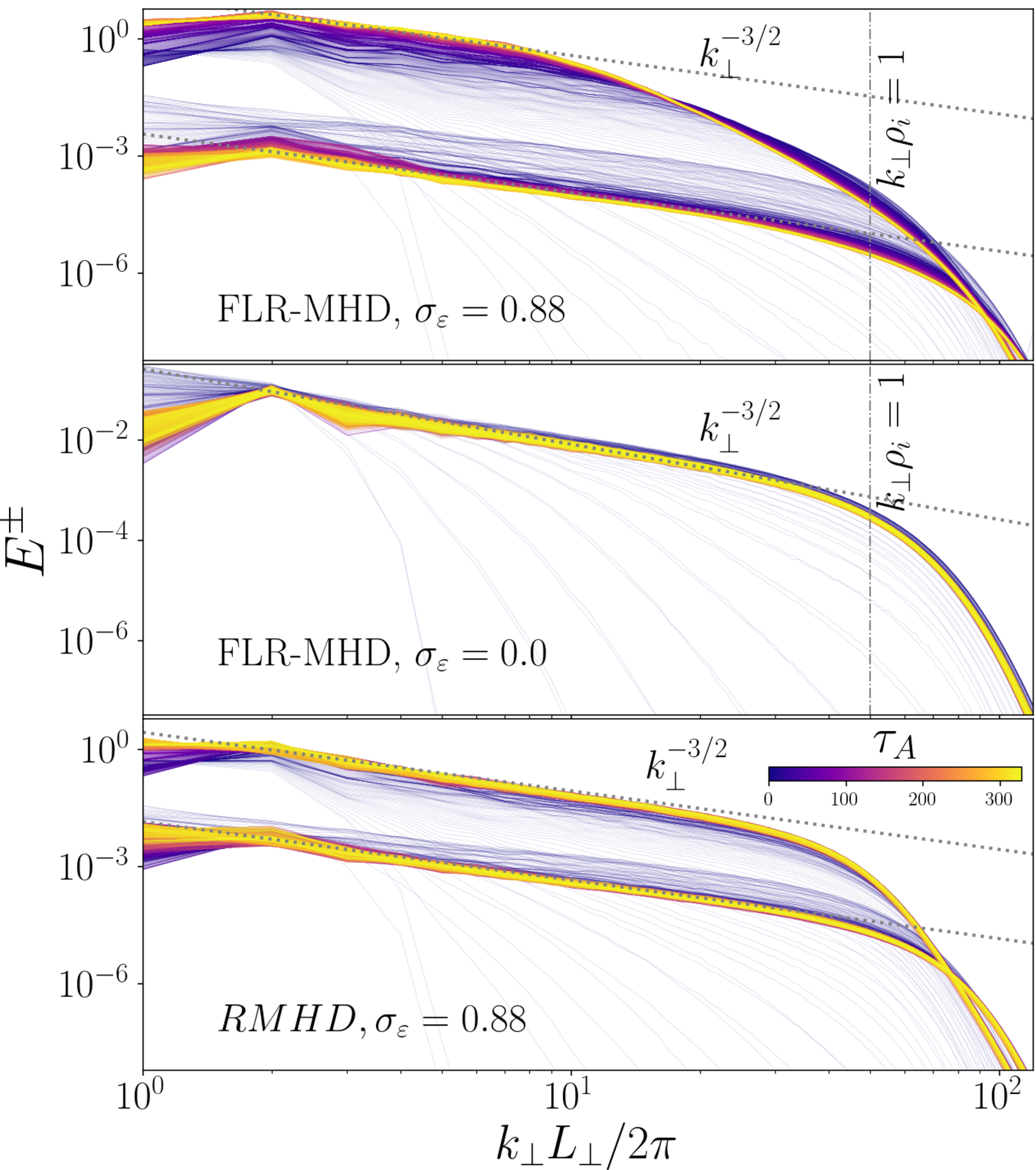}
\caption{Time evolution of the spectra, $E^\pm(k_\perp)$, for the simulations shown in  Fig.~\ref{fig:energy}, comparing imbalanced FLR-MHD (top panel), balanced FLR-MHD (middle panel), and imbalanced RMHD
(bottom panel). Individual spectra are shown at times spaced by $t=0.1\tau_A$, as indicated by the colour. While the spectrum converges rapidly in balanced FLR-MHD and imbalanced RMHD turbulence, the spectra of  imbalanced FLR-MHD turbulence continue to evolve until $t\simeq 200\tau_{A}$, with the break continuously moving to larger scales.   }
\label{fig:time spectra}
\end{figure}

\begin{figure}
\centering
\includegraphics[width=0.64\columnwidth]{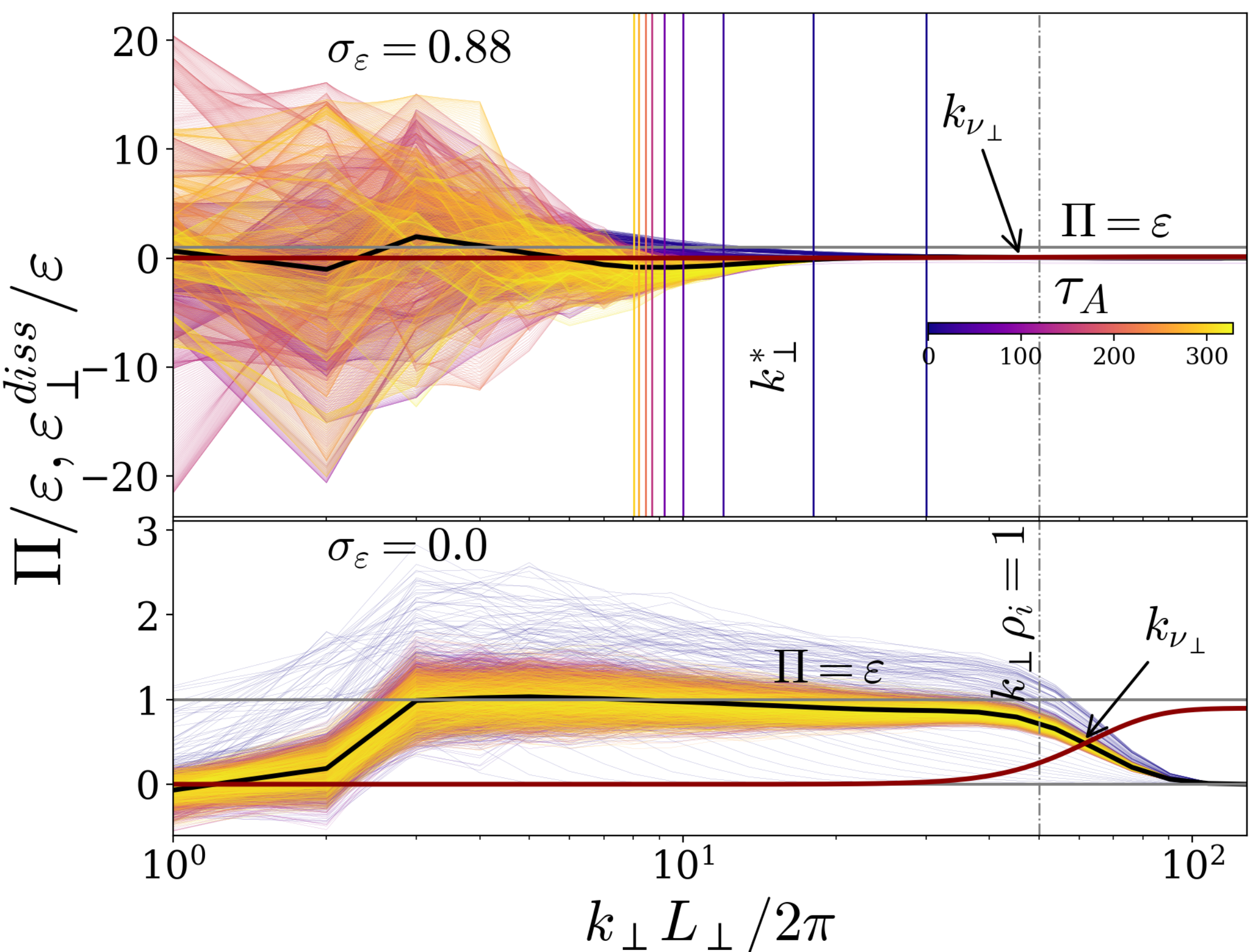}
\caption{Time evolution of the normalised energy flux $\Pi(k_{\perp})/\varepsilon$ for the simulations of   Figs.~\ref{fig:energy}--\ref{fig:time spectra}, comparing imbalanced FLR-MHD (top panel) and balanced FLR-MHD (bottom panel). The  colouring is the same as in Fig.~\ref{fig:time spectra}. While balanced FLR-MHD turbulence shows the expected near-constant flux to small scales (where it is
dissipated), imbalanced FLR-MHD turbulence is characterised by wild fluctuations in $\Pi$ (note different ordinate scale and the position of the grey line at $\Pi=\varepsilon$), which, with time, are increasingly confined to large scales.  The time-dependent wavenumber of the break ($\kstar$) is shown with the coloured vertical lines. \revchng{We also show, with $k_{\nu_\perp}$, the scale at which $\Pi$ is equal to the perpendicular dissipation flux (brown lines) in each simulation. }
The small flux $\Pi\ll \varepsilon$ at smaller scales provides direct evidence for the existence of the helicity barrier.
}
\label{fig:fluxes}
\end{figure}

 Figures~\ref{fig:energy}--\ref{fig:fluxes} show the time evolution of the energy,  dissipation $\varepsilon^{\rm diss}_{\perp,\|}$,  energy spectra $E^{\pm}(k_{\perp})$, and total energy flux $\Pi(k_{\perp})$, 
 comparing imbalanced FLR-MHD  at $\sigf=0.88$ and $\rho_{i} = 0.02 L_{\perp}$\footnote{The saturated state of this simulation is shown with the purple star in \cref{fig:zeroth law}a.}    with balanced FLR-MHD ($\sigf=0$, $\rho_{i} = 0.02 L_{\perp}$) and imbalanced RMHD ($\sigf=0.88$, $\rho_{i} = 0$).
These simulations, which have a resolution of $N_{\perp}=N_{z}=256$, are used as low-resolution seeds (starting at $t\approx18\tau_{A0}$) 
for the recursive resolution refinement that allows us to reach $N_{\perp}=N_{z}=2048$ in Fig.~\ref{fig:spectra}  (the full time evolution is only computationally accessible  at modest resolution).
Let us first describe the balanced FLR-MHD and imbalanced RMHD cases in order to highlight the effect of the helicity barrier. 

The balanced FLR-MHD simulation reaches saturation after a transient phase lasting several $\tau_{A}$,
exhibiting a  $\sim\!k_{\perp}^{-3/2}$ spectrum at large scales (Fig.~\ref{fig:time spectra}) and constant flux of energy to small scales (Fig.~\ref{fig:fluxes}) where it is dissipated with $\edprp\gg\edprl$ (Fig.~\ref{fig:energy}b).
While the transition to KAW turbulence ($\sim\!k_{\perp}^{-2.8}$) at $k_{\perp}\rho_{i}\simeq 1$ is superseded by the dissipation range at $N_{\perp}=256$, it is clearly 
visible in the $N_{\perp}=2048$ spectrum in Fig.~\ref{fig:spectra}.
The imbalanced RMHD simulation is similar, although it is slower to 
saturate, reaching steady state by $\tau_{A}\simeq40$, with a $\sim\!k_{\perp}^{-3/2}$ spectrum in $E^{+}$ and $E^{-}$ (Fig.~\ref{fig:time spectra}) and energy fluxes to small perpendicular scales (not shown).
The larger  saturated energy arises because the cascade time $\tau_{\rm cas}$ is larger in imbalanced turbulence due to 
its slower nonlinear interactions  \citep{Chandran2008,Lithwick2007}, implying 
 $E_{\rm sat}\sim \varepsilon \tau_{\rm cas}$ is larger with fixed  $\varepsilon$.
As $E$ grows, the parallel outer scale $l_{\|0}$  decreases  due to critical balance ($l_{\|0}\sim  L_{\perp}v_{A}/E^{1/2}$), which causes a modest parallel dissipation ($\edprl\simeq0.3 \ed$) observed in RMHD  for the chosen 
parameters (Fig.~\ref{fig:energy}b). This disappears at either lower $\varepsilon$ and/or higher resolution.

Imbalanced FLR-MHD turbulence is markedly different from both its balanced counterpart and imbalanced RMHD turbulence. 
As  noted above, the latter is especially remarkable because  FLR-MHD is identical to RMHD at  $k_{\perp}\rho_{i}\ll1$, and 
$\rho_{i}$ (vertical line in Fig.~\ref{fig:time spectra}) lies only slightly above the resolution cutoff (where perpendicular dissipation dominates) at these parameters. 
After an initial transient phase ($t\lesssim5\tau_{A}$) when its evolution is similar to RMHD, the system forms a
 sharp spectral break at $\kstar$, and the pseudo-stationary phase begins. During
 this phase,  the outer-scale energy in $E^{+}(k_{\perp})$ grows in time, while the spectral break, which lies near $\kstar\rho_{i}\simeq1$ at early times,  
 migrates to larger scales.
That this break is due to the ``helicity barrier''   can be seen directly in the energy flux (Fig.~\ref{fig:fluxes}): as time goes on,
$\Pi(k_{\perp})$ is confined to increasingly large scales (broadly matching $\kstar$, shown with coloured lines), as well as fluctuating wildly compared to balanced turbulence. Clearly, $\Pi \ll \varepsilon$ for $k_{\perp}>\kstar$, which explains the continual increase in $E$ with time during this phase. The subdominant mode's spectrum
 $E^{-}(k_{\perp})$ behaves quite differently to $E^{+}(k_{\perp})$, undergoing a modest decrease at earlier times and saturating well before $E^{+}$. This  implies that the energy imbalance $\sigma_{c}$ increases with time  during the pseudo-stationary phase.
 Interestingly,  the $E^{-}$ cascade appears agnostic to the break in $E^{+}(k_{\perp})$
and  proceeds to small perpendicular scales. This is consistent with the observation that 
the saturated perpendicular energy dissipation seems to approach $\edprp\approx 2\varepsilon^{-}=\varepsilon(1-\sigf)$
in the pseudo-stationary phase (a result that has been confirmed at higher resolution and at other $\sigf$).
This suggests a form of ``flux pinning'' ($\varepsilon^{-}\approx \varepsilon^{+}$),  
whereby the energy flux to small scales is determined by the requirement of a near-balanced KAW cascade (as seen in Fig.~\ref{fig:spectra}), which thus avoids the problems associated with the inverse cascade of helicity. The amplitude of this cascade appears to be
limited by the availability of $\Theta^{-}$ fluctuations arriving from the inertial range.\footnote{\revchng{As seen in \cref{fig:zeroth law}(a) at  small $\nu_{6\perp}$, this result ($\edprp\approx 2\varepsilon^{-}$) 
does not hold in the \emph{saturated} state of FLR-MHD turbulence. This is not 
unexpected, because the parallel dissipation that causes saturation by dissipating energy (see below) can  dissipate  $\Theta^-$ as well as $\Theta^+$, thus reducing the flux of energy to small perpendicular scales  yet further.}}

 The saturation mechanism in imbalanced FLR-MHD  is fundamentally different to the balanced case or to imbalanced RMHD turbulence, because 
  $\Pi(k_{\perp})$ at $k_{\perp}\gtrsim \kstar$  remains limited to $\simeq\!2\varepsilon^{-}$, no matter what the turbulence amplitude. 
 Saturation finally occurs---at $t\approx 200\tau_{A}$ with energy imbalance reaching $\sigma_{c}\approx0.999$ for the FLR-MHD simulation of Figs.~\ref{fig:energy}--\ref{fig:fluxes}---once eddies of perpendicular scale $\kstar$  reach sufficiently large amplitudes and small parallel scales to dissipate through \emph{parallel} hyper-dissipation (Fig.~\ref{fig:energy}). Our simulations indicate that this generation of small parallel scales occurs due to critical balance 
rather than through an independent parallel cascade to 
 small $l_{\|}$ at fixed $k_{\perp}$ (which would imply a $\nu_{6z}$-independent $E_{\rm sat}$).
We can thus estimate the saturation amplitude using  $l_{\|}(k_{\perp})\sim l_{\|0} (L_{\perp}k_{\perp})^{-1/2}$ and 
$Z^{+}_{k_{\perp}}\sim E^{1/2}(L_{\perp}k_{\perp})^{-1/4}$, where $Z^{+}_{k_{\perp}}$ is the typical variation in $\bm{Z}^{+}$ across scale $k_{\perp}^{-1}$
and $l_{\|}(k_{\perp})$ is the corresponding parallel correlation length \citep{Mallet2017a}.
Noting that saturation occurs at $\nu_{6z}l_{\|}(\kstar)^{-6}(Z^{+}_{\kstar})^{2}\sim \varepsilon$ and $l_{\|0}\sim  L_{\perp}v_{A}/E_{\rm sat}^{1/2}$, 
we find \begin{equation}
E_{\rm sat}\sim (\varepsilon/\nu_{6z})^{-1/4} (\kstar L_{\perp})^{-5/8}(v_{A}L_{\perp})^{3/2}.
\end{equation}
  The $\nu_{6z}^{-1/4}$ scaling is approximately satisfied by the simulations in Fig.~\ref{fig:zeroth law} (which all saturate with similar $\kstar$ near the forcing scales), while
the 2-D dissipation spectrum in Fig.~\ref{fig:energy}c   confirms directly that most dissipation occurs  at small $l_{\|}$ on  $\kstar$-scale eddies. Figure \ref{fig:energy}c also shows the critical-balance scaling $l_{\|}\sim k_{\perp}^{-1/2}$ (although note that $k_{z}$ rather than $k_{\|}\sim l_{\|}^{-1}$ is plotted) 
and the finite $\edprp$ at larger $k_{\perp}$ from flux leakage through the barrier. 
As mentioned above, this saturation mechanism is unphysical: by growing to  $\edprl>\edprp$,
the system is trying to break the
 $l_{\|}\ll l_{\perp}$ ordering 
used to derive FLR-MHD. 
Nonetheless, only with this basic understanding of why the system saturates can we evaluate how the helicity barrier 
might evolve in more realistic scenarios.

Finally, it is worth noting an interesting feature of the MHD-scale ($k_{\perp}< \kstar$) turbulence in the pseudo-stationary phase:
even though most of the energy input is unable to be thermalised during this phase, causing 
$E^{+}$ to grow in time, the spectrum remains approximately  $E^{+}(k_{\perp})\sim k_{\perp}^{-3/2}$ for $k_{\perp}< \kstar$ (see \cref{fig:spectra} and \cref{fig:break} inset). This is in 
contrast to  standard  (e.g., hydrodynamic) turbulence with insufficient small-scale dissipation, which usually
forms a thermal spectrum that  is an increasing function of $k_{\perp}$  \citep[see e.g.,][for a study of the truncated Euler equations]{Cichowlas2005,Frisch2008}. 
We speculate that this occurs because, even in the presence of 
a helicity barrier,  $\Theta^{-}$ still nonlinearly cascades to dissipate at small scales (see discussion above),  thus behaving 
in a way similar to RMHD. \revchngtwo{But, for  $k_{\perp}\rho_{i}<1$ where $\bm{Z}^{-}\approx\hat{\bm{z}}\times\nabla_{\perp}\Theta^{-}$, 
the nonlinear cascade of $\Theta^{+}$ is governed by 
 $\Theta^{-}$ because only counter-propagating waves  interact in RMHD. This suggests that $E^{+}$ does not form a thermal spectrum
 because of the $\Theta^{-}$ cascade, although the phenomenology of the cascade in the presence of a barrier remains highly uncertain.
Nonetheless, our simulations robustly show that  both $E^{+}(k_{\perp})$ and $E^{-}(k_{\perp})$ scale as  $\sim\! k_{\perp}^{-3/2}$, as
 observed in the solar wind \citep{Chen2016a,Chen2020}.}

\subsubsection{The ion spectral break}\label{subsub: ion spectral break}

\begin{figure}
\centering
\includegraphics[width=0.7\columnwidth]{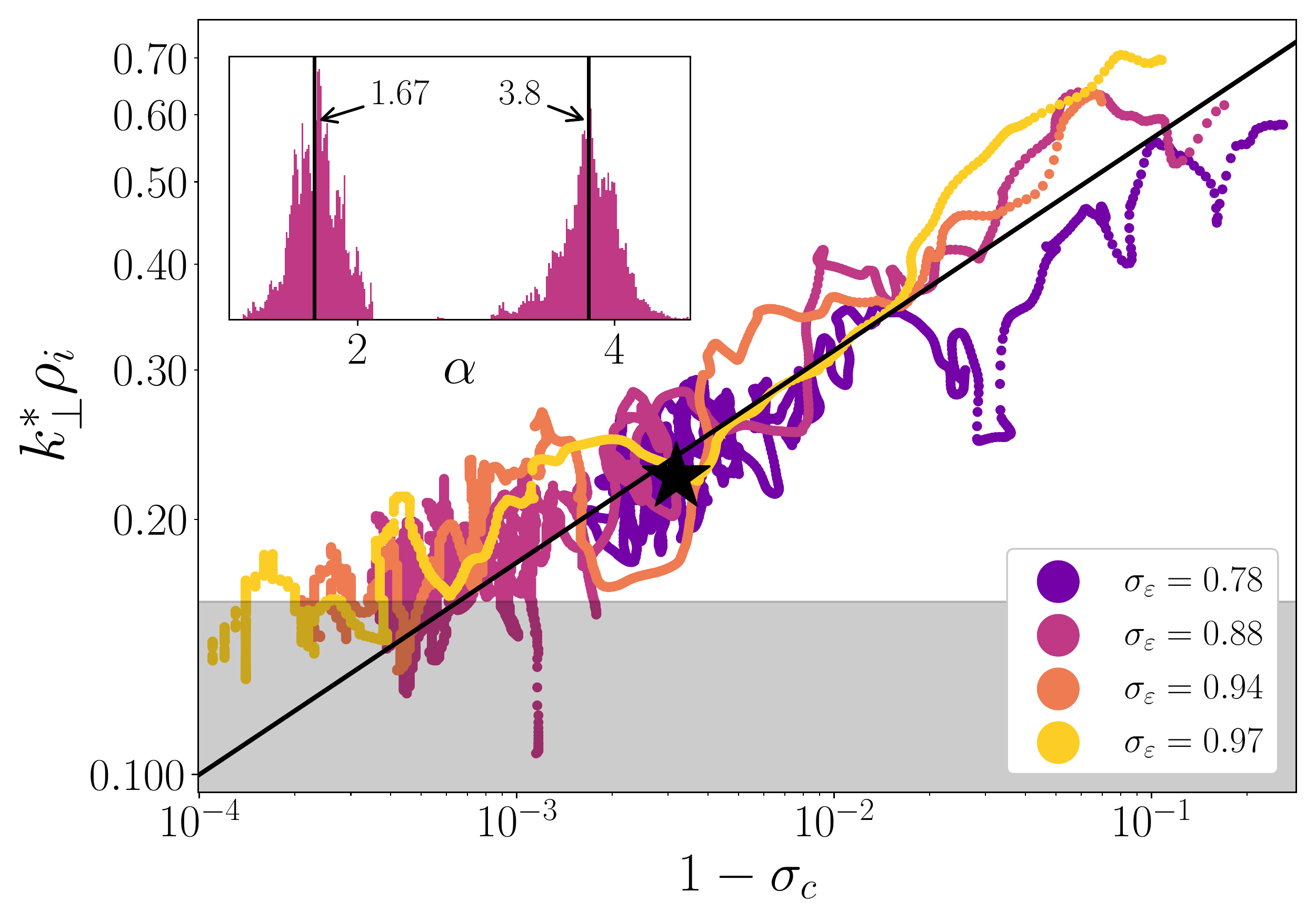}
\caption{Position of the break $\kstar\rho_{i}$ versus the energy imbalance ($1-\sigma_{c}$) for a number of $N_{\perp}=N_{z}=256$ simulations
with different injection imbalances $\sigf$. As $\sigma_{c}(t)$ grows in time due to the 
helicity barrier, there is concurrent decrease in $\kstar$, with no obvious dependence on the helicity injection $\sigf=\varepsilon_{\mathcal{H}}/\varepsilon$
or other time dependence. 
The black line shows the empirical fit \eqref{eq: break fit}, the star shows the fit for Fig.~\ref{fig:spectra}, and the greyed out
region indicates where $\kstar$ gets within a factor of 2 of  the forcing  scale ($\kstar< 4\times 2\pi/L$).   The inset shows a histogram of
the fitted spectral slope $\sim\!k_{\perp}^{-\alpha}$ above and below the break for the $\sigf=0.88$ simulation of Fig.~\ref{fig:time spectra} (the averages are $\langle \alpha\rangle\approx1.67$ above the break and $\langle \alpha\rangle\approx 3.8$  below the break).}
\label{fig:break}
\end{figure}

For testing of the helicity-barrier hypothesis against observations, it is of interest to understand the position 
and spectral slope of the ion-kinetic transition region around $\rho_{i}$ scales. 
Given the unphysical saturation mechanism in our simulations, we hypothesise that the  pseudo-stationary phase is of more relevance to realistic 
space plasmas and that the ``instantaneous'' state of turbulence during that stage can be characterised
by the energy imbalance $\sigma_{c}(t)$ and the injection imbalance $\sigf$, i.e., that the time history of the growth is unimportant.
Figure~\ref{fig:break} shows $\kstar \rho_{i}$ versus imbalance ($1-\sigma_{c}$) for simulations with four different $\sigf$ at $N_{\perp}=N_{z}=256$, 
as well as the spectral slopes ($\sim\!k_{\perp}^{-\alpha}$) above and below $\kstar$ for $\sigf=0.88$ (inset; the values of $\alpha$ are obtained via a broken-power-law fit;  \citealp{astropy}).
We see good correlation of $\kstar$ to $\sigma_{c}(t)$, approximately
\begin{equation}
\kstar \rho_{i} \simeq  (1-\sigma_{c})^{1/4},\label{eq: break fit}
\end{equation}
with little dependence on the injected flux. \revchng{We have also confirmed that the $1/4$-power scaling   
(although not the numerical coefficient) is robust to the order of the parallel hyper-dissipation 
(it also holds if $\nu_{6z}=0$; not shown)}. Spectral slopes in the ion-kinetic transition range ($k_{\perp}> \kstar$) are seen to vary more than
in the MHD range ($k_{\perp}< \kstar$) and are very steep, $\alpha\simeq 4$,
in good agreement with PSP observations \citep{Bowen2020a}.

\section{Discussion}\label{sec: discussion}
Our simulations show a dramatic difference in imbalanced Alfv\'enic turbulence depending on whether or not 
energy can be dissipated at spatial scales above the ion-gyroradius scale. 
If it can, turbulence proceeds in a relatively conventional way, with energy reaching small perpendicular scales where it is thermalised by (hyper)-dissipation.
If it cannot, the helicity barrier blocks the cascade at $\kstar\rho_{i}\lesssim1$, only a small proportion ($\simeq\!2\varepsilon^{-}$) of the energy can reach the smallest perpendicular scales where it would heat electrons, while 
$E^{+}$ grows with time until it becomes so large that modes at $\kstar$ (which itself moves to large scales) dissipate on the parallel viscosity.
The latter effect is unphysical---FLR-MHD is derived by assuming $l_{\|}\gg l_{\perp}$---so the obvious 
question that arises is what would happen in a real plasma such as the solar wind. 
In order for the barrier to lose its importance for large-scale dynamics,  some mechanism must either remove  nearly all of the helicity in the system, thus allowing the
 energy to be channelled into the small-scale KAW cascade,  or significantly dissipate  both 
energy and helicity  at or above the scale of the barrier (as  the parallel dissipation does in FLR-MHD). Presumably a real plasma will find a way to accomplish one of these feats---the
question is which, and what conditions (e.g., fluctuation amplitudes) are required for it to do so. 
A definitive answer will  have to wait  either for observations or for high-resolution six-dimensional kinetic simulations, but we can nonetheless 
speculate on possibilities.

The first possibility is that there exists another perpendicular dissipation mechanism 
that stops the formation of a helicity barrier in the first place. Within the gyrokinetic ordering $l_{\perp}\ll l_{\|}$, because
the low ion-thermal speed at $\beta_{i}\ll1$  implies that Alfv\'enic energy is incapable of heating 
ions significantly
 at any scale, there are in principle three possible such mechanisms: electron Landau damping, electron-inertial effects, and interactions with the compressive cascade. Electron Landau damping is modest 
 at $k_{\perp}\rho_{i}\sim1$ when $1\gg\beta\gg m_{e}/m_{i}$ (e.g., normalised damping of $\simeq\!1\%$ at $\beta\simeq0.1$; see \citealp{Howes2006}), while electron-inertial effects change the equations  only once $k_{\perp}d_{e}\sim1$; so, neither of these 
effects seems capable of  damping substantial helicity or energy. A compressive cascade, although it 
cannot exchange energy with Alfv\'enic motions \citep{Schekochihin2019}, does break helicity conservation around $k_{\perp}\rho_{i}\sim 1$; however, in order to have a significant effect, the compressive  and Alfv\'enic cascades must have similar energy contents, which 
is not generally observed in the solar wind \citep{Bruno2013,Chen2016a,Chen2020}.
Beyond gyrokinetics, cyclotron damping, although perhaps important in the KAW range \citep{Arzamasskiy2019}, requires 
Larmor-frequency fluctuations, a requirement that is difficult to reach for $k_{\perp}\rho_{i}\lesssim1$ fluctuations with $l_{\|}\ll\rho_{i}$. 
Stochastic-ion heating \citep{Chandran2010} is more promising---it can dissipate significant turbulent energy at $k_{\perp}\rho_{i}\sim1$ so long as the
fluctuation amplitude there exceeds a critical threshold $\simeq\!0.2\beta^{1/2}$
---perhaps acting as a dissipation ``switch'' as the amplitude grows.
It is worth noting, however, that if a $\kstar\rho_{i}\lesssim 1$ barrier has formed before significant stochastic heating occurs, 
this reduces the  $\rho_{i}$-scale turbulence amplitude substantially, which also reduces  the heating efficiency. 

If the aforementioned perpendicular dissipation mechanisms fail to dissolve the barrier, it is also possible 
that---even in a real plasma---large-scale energy cannot dissipate, either growing until significant power reaches small parallel scales (of order $d_{i}$ or $\rho_{i}$)  or, in stratified environments such as the solar wind  \citep{Chandran2019}, propagating and growing without dissipation until wave reflection causes the imbalance to decrease enough to allow the cascade to proceed.
In the former case,  various other, non-gyrokinetic dissipation avenues are made available to the plasma; for example,  other cascade channels \citep{Saito2008,Meyrand2012} and ion-cyclotron damping, which could absorb the cascade's energy (much like the parallel hyper-dissipation in 
our simulations).
Interestingly, PSP has observed 
surprisingly high power in $l_{\|}\sim\rho_{i}$ ion-cycloton waves \citep{Huang2020,Bowen2020}, which may be a signature of this mechanism. \revchng{Further, \cite{Duan2021} find a sharp drop in the wavevector anisotropy $l_\|(l_\perp)$ around the break scale, indicating that the dynamics are generating small parallel scales faster than small perpendicular scales. }
Finally, magnetic reconnection may play an important role by enabling nonlocal energy or helicity transfers \citep{Mallet2017,Loureiro2017,Vech2018}; although reconnection
is possible within FLR-MHD, it is necessary to include electron-inertial effects to  capture this physics properly.

\subsection{A critical imbalance?} 
Our theoretical arguments for  helicity-barrier formation, which relied on conservation of energy and helicity in FLR-MHD, suggest that 
a barrier should form with \emph{any} injected  imbalance $\sigf$, but with the constant flux solution failing  at smaller scales for smaller $\sigf$, \emph{viz.,} for $1/ \vph(k_{\perp})<\sigf$.\footnote{Note that this is not incompatible with the observation of Fig.~\ref{fig:break} that the break is independent of $\sigf$: once the 
barrier forms, the system is under no obligation to have a constant flux above the break, meaning the flux can 
go to zero at scales \emph{above} where $\sigf\vph(k_{\perp})\simeq1$.} 
\revchng{ We have confirmed this prediction numerically down to $\sigf\simeq0.3$ (not shown). It is challenging to observe at yet smaller $\sigf$, both because very small scales (compared to $\rho_{i}$) must be resolved, 
and because a large proportion of the energy flux ($\simeq 2\varepsilon^{-}$) is not affected by the barrier, making the resulting growth in 
energy rather slow.} It is, however, clear that non-FLR-MHD effects  will give rise to a critical $\sigf$, below which there is no barrier.
FLR-MHD breaks down at $d_{e}$ scales and helicity is no longer conserved, implying that the helicity 
barrier will likely not form~if \begin{equation}
\sigf\lesssim \frac{1}{\vph(d_{e}^{-1})}\sim \frac{d_{e}}{\rho_{i}}=\frac{Z}{\sqrt{\tau}} \sqrt{\frac{m_{e}}{m_{i}}}\,\beta_{e}^{-1/2}
\end{equation}
(the latter estimates assume $\beta_{e}\gg m_{e}/m_{i}$). The discussion of the previous paragraph also suggests that other effects (e.g., stochastic-ion heating)  would further increase the minimal $\sigf$ for which the barrier forms, by dissipating some helicity or energy.
Measurement of this critical imbalance, along with  improved theoretical understanding of helicity-barrier formation
and dissolution, is left to future  studies.

\subsection{Implications for the solar wind}\label{sub: implications for the solar wind}
The qualitative agreement of our FLR-MHD energy spectra with those observed in the solar wind is highly suggestive. As far as we are aware, no previous numerical simulations have been able to produce similar
double-kinked spectra. 
The position, shape, and cause of the ion-kinetic transition has been a decades-long puzzle with numerous proposed  explanations \citep{Schekochihin2009,Sahraoui2010,Meyrand2012,Lion2016,Voitenko2016,Mallet2017,Woodham2018};  observations show widely varying break positions
and slopes  \citep{Leamon1998} often  followed by a spectral flattening at yet smaller scales \citep{Sahraoui2009,Bowen2020a,Duan2021}.
 In addition,  larger-scale transitions to steeper spectra  
correlate with  higher-amplitude fluctuations, lower $\beta$, higher proton-scale magnetic helicity,
and fast-wind regions \citep{Smith2006,Bruno2014,Vech2018,Zhao2020}, the latter of which is known to be more  imbalanced than the slow wind.
Each of these observations is well explained by the helicity-barrier hypothesis, at least qualitatively: we reproduce double-kinked spectra with the 
first break  at a non-universal scale above $k_{\perp}\rho_{i}=1$, while steeper spectra and  larger-scale breaks
result from the  energy growing in time (Fig.~\ref{fig:time spectra}). It is also worth noting that direct measurement of the 
turbulent energy flux in the solar wind has found the surprising result that 
the  $\bm{Z}^{+}$ flux seems to reverse at large imbalance \citep{Smith2009}; although not fully explained by our 
simulations (\revchng{given that our forcing injects energy only at large scales}), the result does indicate
the presence of a critical imbalance that controls key features of the cascade.
Future observations, combined with more realistic simulations,
 will provide more stringent tests of the theory. 
 
More generally, if the helicity barrier proves to be a robust feature of plasma turbulence,
we see a number of interesting implications. Turbulence is believed to contribute importantly 
to solar-wind heating \citep{Dmitruk2002}, so the requirement that  it  build up significantly in amplitude before being
able to dissipate  may have  consequences for global heliospheric models \citep{Verdini2007,Chandran2009}.
It is also interesting to ask about the plausible relevance to the sudden large-scale field
reversals, or ``switchbacks,'' observed ubiquitously by PSP \citep{Kasper2019,Bale2019}: if switchbacks  form \emph{in-situ} due to wave growth in the expanding plasma \citep{Squire2020}, 
their existence relies on the dominance of growth over dissipation through turbulence. Halting  energy dissipation via the  helicity barrier could thus favor the
development of sharp, large-amplitude structures, as observed. 
Finally, and more generally, the helicity barrier reveals yet another way that weakly collisional plasmas confound standard intuition about their thermodynamics. While Joule found that  water or mercury possess a well-defined 
heat capacity, independent of how the fluid is heated, heating of ions and electrons 
in  a plasma depends not just on bulk properties such as $T_{e}/T_{i}$ or $\beta$ \citep{Howes2008,Kawazura2019}, but also, quite sensitively, 
 on how it is stirred. While the influence of the driving compressibility on heating is already known  \citep{Schekochihin2019,Kawazura2020}, we see that 
 driving imbalance should also have a strong effect, by halting the flow of Alfv\'enic energy to electron scales. 
 The helicity barrier is thus expected to narrow yet further  the range of plasma conditions under which electrons are   heated  preferentially to ions.

%
%

\acknowledgments
{Support for R.M. and J.S.  was
provided by Rutherford Discovery Fellowship RDF-U001804 and Marsden Fund grant UOO1727, which are managed through the Royal Society Te Ap\=arangi. The work of A.A.S. was supported in part by the UK EPSRC programme grant EP/R034737/10. W.D was supported by the US Department of Energy through grant DE-FG02-93ER-54197. The authors wish to acknowledge the use of New Zealand eScience Infrastructure (NeSI) high performance computing facilities and consulting support as part of this research. This work was partially performed using HPC resources from GENCI-CINES (Grant 2019-A0060510871). The authors report no conflict of interest.}

\bibliographystyle{jpp}
\bibliography{bib}

\begin{thebibliography}{76}
\expandafter\ifx\csname natexlab\endcsname\relax\def\natexlab#1{#1}\fi

\bibitem[{Alexakis} \& {Biferale}(2018)]{Alexakis2018}
{\sc {Alexakis}, A. \& {Biferale}, L.} 2018 Cascades and transitions in
  turbulent flows. {\em Phys. Rep.\/} {\bf 767}, 1--101.

\bibitem[{Alexandrova} {\em et~al.\/}(2012){Alexandrova}, {Lacombe},
  {Mangeney}, {Grappin} \& {Maksimovic}]{Alexandrova2012}
{\sc {Alexandrova}, O., {Lacombe}, C., {Mangeney}, A., {Grappin}, R. \&
  {Maksimovic}, M.} 2012 Solar wind turbulent spectrum at plasma kinetic
  scales. {\em Astrophys. J.\/} {\bf 760}, 121.

\bibitem[{Alexandrova} {\em et~al.\/}(2009){Alexandrova}, {Saur}, {Lacombe},
  {Mangeney}, {Mitchell}, {Schwartz} \& {Robert}]{Alexandrova2009}
{\sc {Alexandrova}, O., {Saur}, J., {Lacombe}, C., {Mangeney}, A., {Mitchell},
  J., {Schwartz}, S.~J. \& {Robert}, P.} 2009 Universality of solar-wind
  turbulent spectrum from mhd to electron scales. {\em Phys. Rev. Lett.\/} {\bf
  103}~(16), 165003.

\bibitem[{Arzamasskiy} {\em et~al.\/}(2019){Arzamasskiy}, {Kunz}, {Chandran} \&
  {Quataert}]{Arzamasskiy2019}
{\sc {Arzamasskiy}, L., {Kunz}, M.~W., {Chandran}, B. D.~G. \& {Quataert}, E.}
  2019 Hybrid-kinetic simulations of ion heating in {A}lfv{\'e}nic turbulence.
  {\em Astrophys. J.\/} {\bf 879}~(1), 53.

\bibitem[{Astropy Collaboration}(2013)]{astropy}
{\sc {Astropy Collaboration}} 2013 Astropy: A community python package for
  astronomy. {\em Astron. Astro.\/} {\bf 558}, A33.

\bibitem[{Bale} {\em et~al.\/}(2019){Bale}, {Badman}, {Bonnell} {\em
  et~al.\/}]{Bale2019}
{\sc {Bale}, S.~D., {Badman}, S.~T., {Bonnell}, J.~W. {\em et~al.\/}} 2019
  Highly structured slow solar wind emerging from an equatorial coronal hole.
  {\em Nature\/} {\bf 576}, 237--242.

\bibitem[{Beresnyak}(2014)]{Beresnyak2014}
{\sc {Beresnyak}, A.} 2014 Spectra of strong magnetohydrodynamic turbulence
  from high-resolution simulations. {\em Astrophys. J. Lett.\/} {\bf 784}~(2),
  L20.

\bibitem[{Beresnyak} \& {Lazarian}(2009)]{Beresnyak2009}
{\sc {Beresnyak}, A. \& {Lazarian}, A.} 2009 Structure of stationary strong
  imbalanced turbulence. {\em Astrophys. J.\/} {\bf 702}~(1), 460--471.

\bibitem[Boldyrev(2006)]{Boldyrev2006}
{\sc Boldyrev, S.} 2006 Spectrum of magnetohydrodynamic turbulence. {\em Phys.
  Rev. Lett.\/} {\bf 96}, 115002.

\bibitem[{Boldyrev} {\em et~al.\/}(2013){Boldyrev}, {Horaites}, {Xia} \&
  {Perez}]{Boldyrev2013}
{\sc {Boldyrev}, S., {Horaites}, K., {Xia}, Q. \& {Perez}, J.~C.} 2013 Toward a
  theory of astrophysical plasma turbulence at subproton scales. {\em
  Astrophys. J.\/} {\bf 777}~(1), 41.

\bibitem[{Bowen} {\em et~al.\/}(2020{\natexlab{{\em a\/}}}){Bowen}, {Mallet},
  {Bale} {\em et~al.\/}]{Bowen2020a}
{\sc {Bowen}, T.~A., {Mallet}, A., {Bale}, S.~D. {\em et~al.\/}}
  2020{\natexlab{{\em a\/}}} Constraining ion-scale heating and spectral energy
  transfer in observations of plasma turbulence. {\em Phys. Rev. Lett.\/} {\bf
  125}~(2), 025102.

\bibitem[{Bowen} {\em et~al.\/}(2020{\natexlab{{\em b\/}}}){Bowen}, {Mallet},
  {Huang} {\em et~al.\/}]{Bowen2020}
{\sc {Bowen}, T.~A., {Mallet}, A., {Huang}, J. {\em et~al.\/}}
  2020{\natexlab{{\em b\/}}} Ion-scale electromagnetic waves in the inner
  heliosphere. {\em Astrophys. J. Supp.\/} {\bf 246}~(2), 66.

\bibitem[Bruno \& Carbone(2013)]{Bruno2013}
{\sc Bruno, R. \& Carbone, V.} 2013 The solar wind as a turbulence laboratory.
  {\em Living Rev. Solar Phys.\/} {\bf 10}~(1), 2.

\bibitem[{Bruno} {\em et~al.\/}(2014){Bruno}, {Trenchi} \&
  {Telloni}]{Bruno2014}
{\sc {Bruno}, R., {Trenchi}, L. \& {Telloni}, D.} 2014 Spectral slope variation
  at proton scales from fast to slow solar wind. {\em Astrophys. J. Lett.\/}
  {\bf 793}~(1), L15.

\bibitem[{Chandran}(2008)]{Chandran2008}
{\sc {Chandran}, B. D.~G.} 2008 Strong anisotropic {MHD} turbulence with cross
  helicity. {\em Astrophys. J.\/} {\bf 685}~(1), 646--658.

\bibitem[{Chandran} \& {Hollweg}(2009)]{Chandran2009}
{\sc {Chandran}, B. D.~G. \& {Hollweg}, J.~V.} 2009 {A}lfv{\'e}n wave
  reflection and turbulent heating in the solar wind from 1 solar radius to 1
  {AU}: {A}n analytical treatment. {\em Astrophys. J.\/} {\bf 707}~(2),
  1659--1667.

\bibitem[{Chandran} {\em et~al.\/}(2010){Chandran}, {Li}, {Rogers}, {Quataert}
  \& {Germaschewski}]{Chandran2010}
{\sc {Chandran}, B. D.~G., {Li}, B., {Rogers}, B.~N., {Quataert}, E. \&
  {Germaschewski}, K.} 2010 Perpendicular ion heating by low-frequency
  {A}lfv{\'e}n-wave turbulence in the solar wind. {\em Astrophys. J.\/} {\bf
  720}~(1), 503--515.

\bibitem[{Chandran} \& {Perez}(2019)]{Chandran2019}
{\sc {Chandran}, B. D.~G. \& {Perez}, J.~C.} 2019 Reflection-driven
  magnetohydrodynamic turbulence in the solar atmosphere and solar wind. {\em
  J. Plasma Phys.\/} {\bf 85}~(4), 905850409.

\bibitem[{Chen}(2016)]{Chen2016a}
{\sc {Chen}, C.~H.~K.} 2016 Recent progress in astrophysical plasma turbulence
  from solar wind observations. {\em J. Plasma Phys.\/} {\bf 82}~(6),
  535820602.

\bibitem[Chen {\em et~al.\/}(2020)Chen, Bale, Bonnell {\em et~al.\/}]{Chen2020}
{\sc Chen, C. H.~K., Bale, S.~D., Bonnell, J.~W. {\em et~al.\/}} 2020 The
  evolution and role of solar wind turbulence in the inner heliosphere. {\em
  Astrophys. J. Supp.\/} {\bf 246}~(2), 53.

\bibitem[{Chen} {\em et~al.\/}(2011){Chen}, {Mallet}, {Yousef}, {Schekochihin}
  \& {Horbury}]{Chen2011a}
{\sc {Chen}, C.~H.~K., {Mallet}, A., {Yousef}, T.~A., {Schekochihin}, A.~A. \&
  {Horbury}, T.~S.} 2011 Anisotropy of {A}lfv{\'e}nic turbulence in the solar
  wind and numerical simulations. {\em Mon. Not. R. Astron. Soc.\/} {\bf
  415}~(4), 3219--3226.

\bibitem[{Cho}(2011)]{Cho2011}
{\sc {Cho}, J.} 2011 Magnetic helicity conservation and inverse energy cascade
  in electron magnetohydrodynamic wave packets. {\em Phys. Rev. Lett.\/} {\bf
  106}~(19), 191104.

\bibitem[{Cichowlas} {\em et~al.\/}(2005){Cichowlas}, {Bona{\"\i}ti},
  {Debbasch} \& {Brachet}]{Cichowlas2005}
{\sc {Cichowlas}, C., {Bona{\"\i}ti}, P., {Debbasch}, F. \& {Brachet}, M.} 2005
  Effective dissipation and turbulence in spectrally truncated {E}uler flows.
  {\em Phys. Rev. Lett.\/} {\bf 95}~(26), 264502.

\bibitem[{Dmitruk} {\em et~al.\/}(2002){Dmitruk}, {Matthaeus}, {Milano},
  {Oughton}, {Zank} \& {Mullan}]{Dmitruk2002}
{\sc {Dmitruk}, P., {Matthaeus}, W.~H., {Milano}, L.~J., {Oughton}, S., {Zank},
  G.~P. \& {Mullan}, D.~J.} 2002 Coronal heating distribution due to
  low-frequency, wave-driven turbulence. {\em Astrophys. J.\/} {\bf 575}~(1),
  571--577.

\bibitem[{Dobrowolny} {\em et~al.\/}(1980){Dobrowolny}, {Mangeney} \&
  {Veltri}]{Dobrowolny1980a}
{\sc {Dobrowolny}, M., {Mangeney}, A. \& {Veltri}, P.} 1980 Fully developed
  anisotropic hydromagnetic turbulence in interplanetary space. {\em Phys. Rev.
  Lett.\/} {\bf 45}~(2), 144--147.

\bibitem[{Duan} {\em et~al.\/}(2021){Duan}, {He}, {Bowen}, {Woodham}, {Wang},
  {Chen}, {Mallet} \& {Bale}]{Duan2021}
{\sc {Duan}, D., {He}, J., {Bowen}, T.~A., {Woodham}, L.~D., {Wang}, T.,
  {Chen}, C. H.~K., {Mallet}, A. \& {Bale}, S.~D.} 2021 Anisotropy of
  solar-wind turbulence in the inner heliosphere at kinetic scales: {PSP}
  observations. {\em {arXiv:2102.13294}\/} .

\bibitem[{Elsasser}(1950)]{Elsasser1950}
{\sc {Elsasser}, W.~M.} 1950 The hydromagnetic equations. {\em Phys. Rev.\/}
  {\bf 79}~(1), 183--183.

\bibitem[{Fj{\o}rtoft}(1953)]{Fjortoft1953}
{\sc {Fj{\o}rtoft}, R.} 1953 On the changes in the spectral distribution of
  kinetic energy for twodimensional, nondivergent flow. {\em Tellus\/} {\bf
  5}~(3), 225.

\bibitem[Frisch {\em et~al.\/}(2008)Frisch, Kurien, Pandit, Pauls, Ray, Wirth
  \& Zhu]{Frisch2008}
{\sc Frisch, U., Kurien, S., Pandit, R., Pauls, W., Ray, S.~S., Wirth, A. \&
  Zhu, J.-Z.} 2008 Hyperviscosity, {G}alerkin truncation, and bottlenecks in
  turbulence. {\em Phys. Rev. Lett.\/} {\bf 101}, 144501.

\bibitem[Goldreich \& Sridhar(1995)]{Goldreich1995}
{\sc Goldreich, P. \& Sridhar, S.} 1995 Toward a theory of interstellar
  turbulence. {S}trong {A}lfv{\'e}nic turbulence. {\em Astrophys. J.\/} {\bf
  438}, 763--775.

\bibitem[Howes {\em et~al.\/}(2008)Howes, Cowley, Dorland, Hammett, Quataert \&
  Schekochihin]{Howes2008}
{\sc Howes, G.~G., Cowley, S.~C., Dorland, W., Hammett, G.~W., Quataert, E. \&
  Schekochihin, A.~A.} 2008 A model of turbulence in magnetized plasmas:
  Implications for the dissipation range in the solar wind. {\em J. Geophys.
  Res.: Space Phys.\/} {\bf 113}~(A), A05103.

\bibitem[{Howes} {\em et~al.\/}(2006){Howes}, {Cowley}, {Dorland}, {Hammett},
  {Quataert} \& {Schekochihin}]{Howes2006}
{\sc {Howes}, G.~G., {Cowley}, S.~C., {Dorland}, W., {Hammett}, G.~W.,
  {Quataert}, E. \& {Schekochihin}, A. e.~A.} 2006 Astrophysical gyrokinetics:
  {B}asic equations and linear theory. {\em Astrophys. J.\/} {\bf 651}~(1),
  590--614.

\bibitem[{Howes} {\em et~al.\/}(2008){Howes}, {Dorland}, {Cowley}, {Hammett},
  {Quataert}, {Schekochihin} \& {Tatsuno}]{Howes2008a}
{\sc {Howes}, G.~G., {Dorland}, W., {Cowley}, S.~C., {Hammett}, G.~W.,
  {Quataert}, E., {Schekochihin}, A.~A. \& {Tatsuno}, T.} 2008 Kinetic
  simulations of magnetized turbulence in astrophysical plasmas. {\em Phys.
  Rev. Lett.\/} {\bf 100}~(6), 065004.

\bibitem[{Huang} {\em et~al.\/}(2020){Huang}, {Zhang}, {Sahraoui} {\em
  et~al.\/}]{Huang2020}
{\sc {Huang}, S.~Y., {Zhang}, J., {Sahraoui}, F. {\em et~al.\/}} 2020 Kinetic
  scale slow solar wind turbulence in the inner heliosphere: {C}oexistence of
  kinetic {A}lfv{\'e}n waves and {A}lfv{\'e}n ion cyclotron waves. {\em
  Astrophys. J. Lett.\/} {\bf 897}~(1), L3.

\bibitem[Joule(1850)]{Joule1850}
{\sc Joule, J.~T.} 1850 On the mechanical equivalent of heat. {\em Phil. Trans.
  R. Soc. London\/} {\bf 140}, 61--82.

\bibitem[{Kasper} {\em et~al.\/}(2019){Kasper}, {Bale}, {Belcher} {\em
  et~al.\/}]{Kasper2019}
{\sc {Kasper}, J.~C., {Bale}, S.~D., {Belcher}, J.~W. {\em et~al.\/}} 2019
  Alfv{\'e}nic velocity spikes and rotational flows in the near-sun solar wind.
  {\em Nature\/} {\bf 576}, 228--231.

\bibitem[{Kawazura} {\em et~al.\/}(2019){Kawazura}, {Barnes} \&
  {Schekochihin}]{Kawazura2019}
{\sc {Kawazura}, Y., {Barnes}, M. \& {Schekochihin}, A.~A.} 2019 Thermal
  disequilibration of ions and electrons by collisionless plasma turbulence.
  {\em Proc. Nat. Acc. Sci.\/} {\bf 116}~(3), 771--776.

\bibitem[{Kawazura} {\em et~al.\/}(2020){Kawazura}, {Schekochihin}, {Barnes},
  {TenBarge}, {Tong}, {Klein} \& {Dorland}]{Kawazura2020}
{\sc {Kawazura}, Y., {Schekochihin}, A.~A., {Barnes}, M., {TenBarge}, J.~M.,
  {Tong}, Y., {Klein}, K.~G. \& {Dorland}, W.} 2020 Ion versus electron heating
  in compressively driven astrophysical gyrokinetic turbulence. {\em Phys. Rev.
  X\/} {\bf 10}~(4), 041050.

\bibitem[{Kim} \& {Cho}(2015)]{Kim2015}
{\sc {Kim}, H. \& {Cho}, J.} 2015 Inverse cascade in imbalanced electron
  magnetohydrodynamic turbulence. {\em Astrophys. J.\/} {\bf 801}~(2), 75.

\bibitem[{Leamon} {\em et~al.\/}(1998){Leamon}, {Smith}, {Ness}, {Matthaeus} \&
  {Wong}]{Leamon1998}
{\sc {Leamon}, R.~J., {Smith}, C.~W., {Ness}, N.~F., {Matthaeus}, W.~H. \&
  {Wong}, H.~K.} 1998 {Observational constraints on the dynamics of the
  interplanetary magnetic field dissipation range}. {\em J. Geophys. Res\/}
  {\bf 103}~(A3), 4775--4788.

\bibitem[{Lion} {\em et~al.\/}(2016){Lion}, {Alexandrova} \&
  {Zaslavsky}]{Lion2016}
{\sc {Lion}, S., {Alexandrova}, O. \& {Zaslavsky}, A.} 2016 Coherent events and
  spectral shape at ion kinetic scales in the fast solar wind turbulence. {\em
  Astrophys. J.\/} {\bf 824}~(1), 47.

\bibitem[Lithwick {\em et~al.\/}(2007)Lithwick, Goldreich \&
  Sridhar]{Lithwick2007}
{\sc Lithwick, Y., Goldreich, P. \& Sridhar, S.} 2007 Imbalanced strong {MHD}
  turbulence. {\em Astrophys. J.\/} {\bf 655}~(1), 269--274.

\bibitem[{Loureiro} \& {Boldyrev}(2017)]{Loureiro2017}
{\sc {Loureiro}, N.~F. \& {Boldyrev}, S.} 2017 Collisionless reconnection in
  magnetohydrodynamic and kinetic turbulence. {\em Astrophys. J.\/} {\bf
  850}~(2), 182.

\bibitem[{Mallet} \& {Schekochihin}(2017)]{Mallet2017a}
{\sc {Mallet}, A. \& {Schekochihin}, A.~A.} 2017 A statistical model of
  three-dimensional anisotropy and intermittency in strong alfv{\'e}nic
  turbulence. {\em Mon. Not. R. Astron. Soc.\/} {\bf 466}~(4), 3918--3927.

\bibitem[{Mallet} {\em et~al.\/}(2017){Mallet}, {Schekochihin} \&
  {Chandran}]{Mallet2017}
{\sc {Mallet}, A., {Schekochihin}, A.~A. \& {Chandran}, B. D.~G.} 2017
  Disruption of {A}lfv{\'e}nic turbulence by magnetic reconnection in a
  collisionless plasma. {\em J. Plasma Phys.\/} {\bf 83}~(6), 905830609.

\bibitem[Maron \& Goldreich(2001)]{Maron2001}
{\sc Maron, J. \& Goldreich, P.} 2001 Simulations of incompressible
  magnetohydrodynamic turbulence. {\em Astrophys. J.\/} {\bf 554}~(2),
  1175--1196.

\bibitem[{McManus} {\em et~al.\/}(2020){McManus}, {Bowen}, {Mallet} {\em
  et~al.\/}]{McManus2020}
{\sc {McManus}, M.~D., {Bowen}, T.~A., {Mallet}, A. {\em et~al.\/}} 2020 Cross
  helicity reversals in magnetic switchbacks. {\em Astrophys. J. Supp.\/} {\bf
  246}~(2), 67.

\bibitem[Meyrand \& Galtier(2012)]{Meyrand2012}
{\sc Meyrand, R. \& Galtier, S.} 2012 Spontaneous chiral symmetry breaking of
  {Hall} magnetohydrodynamic turbulence. {\em Phys. Rev. Lett.\/} {\bf 109},
  194501.

\bibitem[{Meyrand} {\em et~al.\/}(2019){Meyrand}, {Kanekar}, {Dorland} \&
  {Schekochihin}]{Meyrand2019}
{\sc {Meyrand}, R., {Kanekar}, A., {Dorland}, W. \& {Schekochihin}, A.~A.} 2019
  Fluidization of collisionless plasma turbulence. {\em Proc. Nat. Acc. Sci.\/}
  {\bf 116}~(4), 1185--1194.

\bibitem[{Milanese} {\em et~al.\/}(2020){Milanese}, {Loureiro}, {Daschner} \&
  {Boldyrev}]{Milanese2020}
{\sc {Milanese}, L.~M., {Loureiro}, N.~F., {Daschner}, M. \& {Boldyrev}, S.}
  2020 Dynamic phase alignment in inertial {A}lfv{\'e}n turbulence. {\em Phys.
  Rev. Lett.\/} {\bf 125}~(26), 265101.

\bibitem[{Miloshevich} {\em et~al.\/}(2020){Miloshevich}, {Laveder}, {Passot}
  \& {Sulem}]{Miloshevich2020}
{\sc {Miloshevich}, G., {Laveder}, D., {Passot}, T. \& {Sulem}, P.-L.} 2020
  Inverse cascade and magnetic vortices in kinetic {A}lfv{\'e}n-wave
  turbulence. {\em arXiv:2007.06976\/} .

\bibitem[{Oughton} {\em et~al.\/}(1994){Oughton}, {Priest} \&
  {Matthaeus}]{Oughton1994}
{\sc {Oughton}, S., {Priest}, E.~R. \& {Matthaeus}, W.~H.} 1994 The influence
  of a mean magnetic field on three-dimensional magnetohydrodynamic turbulence.
  {\em J. Fluid Mech.\/} {\bf 280}, 95--117.

\bibitem[{Passot} {\em et~al.\/}(2018){Passot}, {Sulem} \& {Tassi}]{Passot2018}
{\sc {Passot}, T., {Sulem}, P.~L. \& {Tassi}, E.} 2018 {Gyrofluid modeling and
  phenomenology of low-{\ensuremath{\beta}}$_{e}$ Alfv{\'e}n wave turbulence}.
  {\em Phys. Plasmas\/} {\bf 25}~(4), 042107.

\bibitem[{Pearson} {\em et~al.\/}(2004){Pearson}, {Yousef}, {Haugen},
  {Brandenburg} \& {Krogstad}]{Pearson2004}
{\sc {Pearson}, B.~R., {Yousef}, T.~A., {Haugen}, N. E.~L., {Brandenburg}, A.
  \& {Krogstad}, P.-{\r{A}}.} 2004 Delayed correlation between turbulent energy
  injection and dissipation. {\em Phys. Rev. E\/} {\bf 70}~(5), 056301.

\bibitem[{Perez} \& {Boldyrev}(2009)]{Perez2009}
{\sc {Perez}, J.~C. \& {Boldyrev}, S.} 2009 Role of cross-helicity in
  magnetohydrodynamic turbulence. {\em Phys. Rev. Lett.\/} .

\bibitem[Perez {\em et~al.\/}(2012)Perez, Mason, Boldyrev \&
  Cattaneo]{Perez2012}
{\sc Perez, J.~C., Mason, J., Boldyrev, S. \& Cattaneo, F.} 2012 On the energy
  spectrum of strong magnetohydrodynamic turbulence. {\em Phys. Rev. X\/} {\bf
  2}, 041005.

\bibitem[Pouquet {\em et~al.\/}(2020)Pouquet, Stawarz \&
  Rosenberg]{Pouquet2020}
{\sc Pouquet, A., Stawarz, J.~E. \& Rosenberg, D.} 2020 Coupling large eddies
  and waves in turbulence: Case study of magnetic helicity at the ion inertial
  scale. {\em Atmosphere\/} {\bf 11}~(2), 203.

\bibitem[Sahraoui {\em et~al.\/}(2010)Sahraoui, Goldstein, Belmont, Canu \&
  Rezeau]{Sahraoui2010}
{\sc Sahraoui, F., Goldstein, M.~L., Belmont, G., Canu, P. \& Rezeau, L.} 2010
  Three dimensional anisotropic $k$ spectra of turbulence at subproton scales
  in the solar wind. {\em Phys. Rev. Lett.\/} {\bf 105}, 131101.

\bibitem[{Sahraoui} {\em et~al.\/}(2009){Sahraoui}, {Goldstein}, {Robert} \&
  {Khotyaintsev}]{Sahraoui2009}
{\sc {Sahraoui}, F., {Goldstein}, M.~L., {Robert}, P. \& {Khotyaintsev}, Y.~V.}
  2009 Evidence of a cascade and dissipation of solar-wind turbulence at the
  electron gyroscale. {\em Phys. Rev. Lett.\/} {\bf 102}~(23), 231102.

\bibitem[{Saito} {\em et~al.\/}(2008){Saito}, {Gary}, {Li} \&
  {Narita}]{Saito2008}
{\sc {Saito}, S., {Gary}, S.~P., {Li}, H. \& {Narita}, Y.} 2008 Whistler
  turbulence: Particle-in-cell simulations. {\em Phys. Plasmas\/} {\bf
  15}~(10), 102305.

\bibitem[{Schekochihin}(2020)]{Schekochihin2020}
{\sc {Schekochihin}, A.~A.} 2020 {MHD} turbulence: A biased review. {\em
  arXiv:2010.00699\/} .

\bibitem[Schekochihin {\em et~al.\/}(2009)Schekochihin, Cowley, Dorland,
  Hammett, Howes, Quataert \& Tatsuno]{Schekochihin2009}
{\sc Schekochihin, A.~A., Cowley, S.~C., Dorland, W., Hammett, G.~W., Howes,
  G.~G., Quataert, E. \& Tatsuno, T.} 2009 Astrophysical gyrokinetics: Kinetic
  and fluid turbulent cascades in magnetized weakly collisional plasmas. {\em
  Astrophys. J. Supp.\/} {\bf 182}~(1), 310.

\bibitem[{Schekochihin} {\em et~al.\/}(2019){Schekochihin}, {Kawazura} \&
  {Barnes}]{Schekochihin2019}
{\sc {Schekochihin}, A.~A., {Kawazura}, Y. \& {Barnes}, M.~A.} 2019 Constraints
  on ion versus electron heating by plasma turbulence at low beta. {\em J.
  Plasma Phys.\/} {\bf 85}~(3), 905850303.

\bibitem[{Smith} {\em et~al.\/}(2006){Smith}, {Hamilton}, {Vasquez} \&
  {Leamon}]{Smith2006}
{\sc {Smith}, C.~W., {Hamilton}, K., {Vasquez}, B.~J. \& {Leamon}, R.~J.} 2006
  Dependence of the dissipation range spectrum of interplanetary magnetic
  fluctuations on the rate of energy cascade. {\em Astrophys. J. Lett.\/} {\bf
  645}~(1), L85--L88.

\bibitem[{Smith} {\em et~al.\/}(2009){Smith}, {Stawarz}, {Vasquez}, {Forman} \&
  {MacBride}]{Smith2009}
{\sc {Smith}, C.~W., {Stawarz}, J.~E., {Vasquez}, B.~J., {Forman}, M.~A. \&
  {MacBride}, B.~T.} 2009 Turbulent cascade at 1 au in high cross-helicity
  flows. {\em Phys. Rev. Lett.\/} {\bf 103}~(20), 201101.

\bibitem[{Squire} {\em et~al.\/}(2020){Squire}, {Chandran} \&
  {Meyrand}]{Squire2020}
{\sc {Squire}, J., {Chandran}, B.~D.~G. \& {Meyrand}, R.} 2020 In-situ
  switchback formation in the expanding solar wind. {\em Astrophys. J. Lett.\/}
  {\bf 891}~(1), L2.

\bibitem[{Strauss}(1976)]{Strauss1976}
{\sc {Strauss}, H.~R.} 1976 Nonlinear, three-dimensional magnetohydrodynamics
  of noncircular tokamaks. {\em Phys. Fluids\/} {\bf 19}~(1), 134--140.

\bibitem[Teaca {\em et~al.\/}(2009)Teaca, Verma, Knaepen \& Carati]{Teaca2009}
{\sc Teaca, B., Verma, M.~K., Knaepen, B. \& Carati, D.} 2009 Energy transfer
  in anisotropic magnetohydrodynamic turbulence. {\em Phys. Rev. E\/} {\bf 79},
  046312.

\bibitem[{Vech} {\em et~al.\/}(2018){Vech}, {Mallet}, {Klein} \&
  {Kasper}]{Vech2018}
{\sc {Vech}, D., {Mallet}, A., {Klein}, K.~G. \& {Kasper}, J.~C.} 2018 Magnetic
  reconnection may control the ion-scale spectral break of solar wind
  turbulence. {\em Astrophys. J. Lett.\/} {\bf 855}~(2), L27.

\bibitem[Velli(1993)]{Velli1993}
{\sc Velli, M.} 1993 On the propagation of ideal, linear {A}lfv{\'e}n waves in
  radially stratified stellar atmospheres and winds. {\em Astron. Astro.\/}
  {\bf 270}, 304--314.

\bibitem[{Verdini} \& {Velli}(2007)]{Verdini2007}
{\sc {Verdini}, A. \& {Velli}, M.} 2007 {A}lfv{\'e}n waves and turbulence in
  the solar atmosphere and solar wind. {\em Astrophys. J.\/} {\bf 662}~(1),
  669--676.

\bibitem[{Voitenko} \& {De Keyser}(2016)]{Voitenko2016}
{\sc {Voitenko}, Y. \& {De Keyser}, J.} 2016 {MHD}-kinetic transition in
  imbalanced {A}lfv{\'e}nic turbulence. {\em Astrophys. J. Lett.\/} {\bf
  832}~(2), L20.

\bibitem[Williamson(1980)]{Williamson1980}
{\sc Williamson, J.} 1980 Low-storage {Runge-Kutta} schemes. {\em J. Comp.
  Phys.\/} {\bf 35}~(1), 48 -- 56.

\bibitem[Woodham {\em et~al.\/}(2018)Woodham, Wicks, Verscharen \&
  Owen]{Woodham2018}
{\sc Woodham, L.~D., Wicks, R.~T., Verscharen, D. \& Owen, C.~J.} 2018 The role
  of proton cyclotron resonance as a dissipation mechanism in solar wind
  turbulence: A statistical study at ion-kinetic scales. {\em Astrophys. J.\/}
  {\bf 856}~(1), 49.

\bibitem[{Zhao} {\em et~al.\/}(2020){Zhao}, {Lin}, {Wang}, {Wu}, {Feng}, {Liu},
  {Zhao} \& {Li}]{Zhao2020}
{\sc {Zhao}, G.~Q., {Lin}, Y., {Wang}, X.~Y., {Wu}, D.~J., {Feng}, H.~Q.,
  {Liu}, Q., {Zhao}, A. \& {Li}, H.~B.} 2020 Observational evidence for solar
  wind proton heating by ion-scale turbulence. {\em Geophysical Research
  Letters\/} {\bf 47}~(18), e89720.

\bibitem[{Zocco} \& {Schekochihin}(2011)]{Zocco2011}
{\sc {Zocco}, A. \& {Schekochihin}, A.~A.} 2011 Reduced fluid-kinetic equations
  for low-frequency dynamics, magnetic reconnection, and electron heating in
  low-beta plasmas. {\em Phys. Plasmas\/} {\bf 18}~(10), 102309.

\end{thebibliography}

\end{document}